\documentclass[12pt]{article}

\usepackage{graphicx}
\usepackage{color}
\usepackage{bm}

\def\eq#1{(\ref{#1})}

\newcommand{\ket}[1]{|{#1}\rangle}

\def\beq{\begin{equation}}
\def\eeq{\end{equation}}
\def\beqa{\begin{eqnarray}}
\def\eeqa{\end{eqnarray}}
\def\bet{\begin{tabular}}
\def\eet{\end{tabular}}

\newcommand{\ex}[1]{{\rm e}^{#1}} \def\ii{{\rm i}}
\newcommand{\Tr}{{\rm Tr}}
\newcommand{\sect}[1]{\setcounter{equation}{0}\section{#1}}

\renewcommand{\a}{\alpha}

\newcommand{\e}{\epsilon}

\def\one{{\hbox{ 1\kern-.8mm l}}}

\usepackage[centertags]{amsmath}

\usepackage{amsfonts} 
\usepackage{amssymb} 
\usepackage{amsthm}

\begin{document}

\begin{titlepage}

\setcounter{page}{0}

\begin{flushright}
{DFTT 16/2007}\\
{QMUL-PH-07-17}
\end{flushright}

\vspace{0.6cm}

\begin{center}
{\Large \bf New twist field couplings from the \\ 
     partition function for multiply wrapped D-branes.} \\ 

\vskip 0.8cm

{\bf Dario Du\`o and Rodolfo Russo}\\
{\sl 
Centre for Research in String Theory \\ Department of Physics\\
Queen Mary, University of London\\
Mile End Road, London, E1 4NS,
United Kingdom}\\

\vskip .3cm

{\bf Stefano Sciuto}\\
{\sl Dipartimento di Fisica Teorica, Universit\`a di Torino}\\
 and {\sl  INFN, Sezione di Torino}\\
{\sl Via P. Giuria 1, I-10125 Torino, Italy}\\

\vskip 1.2cm

\end{center}

\begin{abstract}

We consider toroidal compactifications of bosonic string theory with
particular regard to the phases (cocycles) necessary for a consistent
definition of the vertex operators, the boundary states and the
T-duality rules. We use these ingredients to compute the planar
multi-loop partition function describing the interaction among
magnetized or intersecting D-branes, also in presence of open string
moduli.  It turns out that unitarity in the open string channel
crucially depends on the presence of the cocycles. We then focus on
the 2-loop case and study the degeneration limit where this partition
function is directly related to the tree-level 3-point correlators
between twist fields. These correlators represent the main ingredient
in the computation of Yukawa couplings and other terms in the
effective action for D-brane phenomenological models. By factorizing
the 2-loop partition function we are able to compute the 3-point
couplings for abelian twist fields on generic non-factorized tori,
thus generalizing previous expressions valid for the 2-torus.

\end{abstract}

\vfill

\end{titlepage}

\sect{Introduction}\label{Intro}

The relation between the modular and the unitarity properties of open
string amplitudes have played a crucial role in deepening our
understanding of string theory. For instance,
Lovelace~\cite{Lovelace:1971fa} studied the modular transformation of
1-loop non-planar open string amplitudes. By requiring that these
amplitudes do not contain cuts, he discovered the critical dimension
of bosonic string theory. More than twenty years later
Polchinski~\cite{Polchinski:1995mt} used the same modular
transformation between the open and the closed string channels on the
1-loop partition function with Neumann and Dirichlet boundary
conditions. By doing so he was able to show that D-branes are actually
dynamical objects that have gravitational couplings with closed
strings.

\begin{figure}
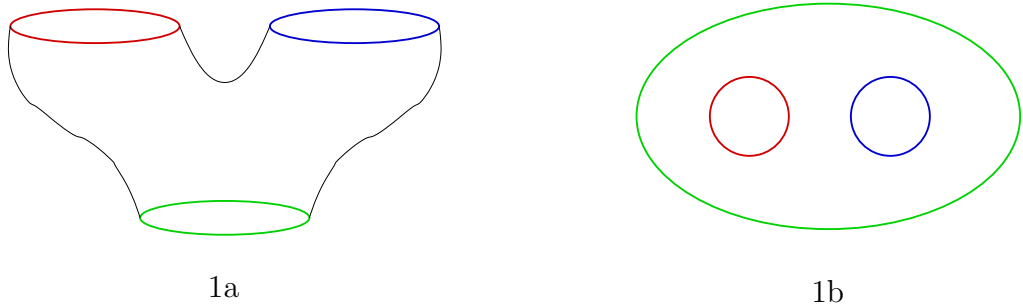

  \begin{center}
\input g2.pstex_t
  \end{center}
 \caption{\label{2l}The twisted open string partition function with
  three borders and no handles ({\em i.e.} on each boundary
     different left/right gluing conditions are imposed). In~\ref{2l}a
    the amplitude is depicted in the closed string channel, while
   in~\ref{2l}b it is modular transformed and the open string
  propagation is manifest.}
\end{figure}

The same interplay between modular transformations and unitarity
properties exists also for higher loop amplitudes. For instance, the
open string diagram for the 2-loop planar partition function is a
Riemann surface with three boundaries and no handles.  This surface
can be described either in the closed string channel as in
Fig.~\ref{2l}a, or in the open channel as in Fig.~\ref{2l}b. In the
first case, by unitarity, one should be able to decompose the result
into a $3$-vertex among closed strings and three boundary states.  In
the open string parametrization, on the other hand, the same amplitude
should factorize into three open strings propagators and two $3$-point
vertices among open strings. This double description of the 2-loop
partition function is particularly interesting when different
left/right gluing conditions are imposed on the various boundaries.
In~\cite{Russo:2007tc} the bosonic contribution to the twisted
$g$-loop partition function was computed by using the closed string
description and then was modular transformed in the open string
channel. It was also shown that the factorization of the 2-loop
diagram provides an effective strategy to compute the couplings among
twists fields ($\sigma_\epsilon$), which is alternative to the
stress-energy tensor technique of~\cite{Dixon:1986qv}. From the
Conformal Field Theory (CFT) point of view, these $\sigma_\epsilon$'s
are operators implementing a change in the boundary conditions for the
(complexified) bosonic coordinates.  The boundary conditions induced
by the $\sigma_\epsilon$'s are appropriate to describe open strings
stretched between D-branes with constant magnetic
fields~\cite{Abouelsaood:1986gd} or D-branes at
angles~\cite{Berkooz:1996km}. This kind of open strings is one of the
main ingredients in D-brane model building (for a recent review
see~\cite{Marchesano:2007de} and the references therein). The
$3$-twist field correlators mentioned above provide the non-trivial
part of the Yukawa couplings\footnote{Actually the twist field
  couplings play the same role also in phenomenologically interesting
  compactifications of Heterotic string
  theory~\cite{Dixon:1986qv,Burwick:1990tu,Erler:1992gt,Stieberger:1992bj,Stieberger:1992vb}
}~\cite{Cremades:2003qj,Cvetic:2003ch,Abel:2003vv,Abel:2003yx,Lust:2004cx,Russo:2007tc}
and of other terms in the effective action generated by stringy
instantons~\cite{Blumenhagen:2006xt,Ibanez:2006da}.

In this paper we generalize the results of~\cite{Russo:2007tc} in
various ways. We compute the bosonic contribution to the planar
$g$-loop partition function for open strings on generic tori with both
multiply wound D-branes and non-zero Wilson lines or open string
moduli. Then we use this result to derive an explicit expression for
the 3-twist field correlators valid beyond the case of 2-torus.
However, in our computations, we still keep a technical assumption:
the monodromy matrices characterizing the open strings stretched
between the D-branes must commute, see Eq.~\eq{coms}. This is always
the case when the gauge field strengths on the various branes are
themselves commuting or parallel. However this more stringent
condition is not necessary and we are able to compute the partition
function also for some setups involving oblique fluxes (according to
the nomenclature of~\cite{Bianchi:2005yz}). On the other hand the
assumption~\eq{coms} will restrict our final results on the twist field
couplings to the abelian case. The D-brane configurations studied in
this paper can be promoted to supersymmetry preserving setups, once
they are embedded in superstring theory. This is achieved simply by
imposing some constraints on the magnetic fluxes to satisfy the D-term
and the F-term conditions. Moreover the latter ones are often
automatically implied by our hypothesis~\eq{coms}, since it ensures
that there is a particular complex basis for the torus where almost
all magnetic fluxes are described by $(1,1)$-forms.

As it was done in~\cite{Russo:2007tc}, we compute the twisted
partition function in the closed string channel by using the operator
formalism (see~\cite{Alvarez-Gaume:1988bg,DiVecchia:1988cy} for
detailed discussion). This approach requires particular care in
dealing with some phase factors present in the definition of the
string vertex operators and boundary states.
The origin of these phases is
well-known: in toroidal compactifications the logarithmic branch cut of
the bosonic Green function can sometime become visible and it is
necessary to compensate for this by adding to the string vertices a
phase known as cocycle (see for instance the paragraph ``A
technicality'' in Sect.  8.2 of~\cite{Polchinski:1998rq}). All cocycle
factors were ignored in~\cite{Russo:2007tc}, but this had no visible
effects because, as we will show, these phases were not crucial in the
particular examples considered there. However, in general, the
partition function is unitary, in the open string channel, only when
all phases have been taken into account. This comes as no surprise
because the partition function of Fig.~\ref{2l}a contains a sum over
all possible $3$-string vertices and the effect of the cocycles
becomes clearly visible since they yield relative phases between
different terms. The presence of these cocycles has some consequences
also on the precise formulation of the T-duality transformations. In
fact these transformations should preserve both the spectrum and the
interactions, including, in the latter case, the phases needed for
ensuring the locality of the interactions. This does not happen with
the naive version of the T-duality rules usually written, and we show
that it is necessary to introduce some cocycle phases also in the
T-duality transformation to recover a consistent picture.

Then we study the degeneration limit of the $g=2$ partition function
in the open string channel, see Fig.~\ref{2l}b, and derive an explicit
form for the 3-twist field correlators on arbitrary tori. This
computation, in presence of multiply wrapped D-branes, provides a
concrete example showing that the cocycle phases are crucial for
unitarity already at the level of the 2-loop partition function.
Our result on the 3-twist field couplings generalizes previous
expressions valid for configurations that are completely factorized in
a product of
2-tori~\cite{Cremades:2003qj,Cvetic:2003ch,Abel:2003vv,Abel:2003yx,Lust:2004cx,Russo:2007tc},
which requires that both the background geometry and the D-brane gauge
field strengths are non-trivial only along the same orthogonal $T^2$'s
and all ``off-diagonal'' entries are put to zero. Even in presence of
commuting fluxes this is a very particular situation, where the
so-called classical contribution to the 3-twist field correlators is
written in terms of Jacobi Theta-functions. In general, higher genus
Theta-functions appear and so the Yukawa couplings that arise in
non-factorized D-brane models will have a more complicated structure
and a richer moduli dependence than those derived so far for the
$T^2$. However, we still find the same separation between K\"ahler and
complex structure moduli and only the latter ones enter in the
expressions for the classical contribution (this in the magnetized
setup, of course the roles are switched in the language of D-branes at
angles).

The structure of the paper is the following. In Section~\ref{t2n} we
briefly review the main features of the string dynamics in toroidal
compactifications. We pay particular care to the cocycle factors in
the definition of the vertices and of the boundary states. We show
that these phases enter in a non-trivial way also in the T-duality
transformation of the states with non-zero Kaluza-Klein or winding
numbers. In Section~\ref{SectionAmplitude} we revisit the computation
of the twisted partition function discussed in~\cite{Russo:2007tc} and
include the effect of the Wilson lines and the D-brane multiple
wrappings.  The only hypothesis we still take is to restrict ourselves
to the case of commuting monodromy matrices (see Eq.~\eq{coms}). In
Section~\ref{tft2d} we focus on the $g=2$ case and study the unitarity
properties of the result. By doing so we derive the general expression
for the 3-point correlators between abelian twist fields. Finally some
technicalities are left to the Appendix.

\sect{Toroidal compactifications}\label{t2n}

\subsection{Some properties of $T^{2d}$}

A $2d$-dimensional real torus $T^{2d}$ is defined by a collection of
$2d$ vectors $a_M$ in the Euclidean space $\mathbb{R}^{2d}$: $T^{2d}$
is simply $\mathbb{R}^{2d}$ modulo the identification with integer
shifts along the $a_M$'s
\begin{equation}\label{lattice}
x \equiv x+2\pi \sqrt{\alpha'}\sum_{M=1}^{2d}c^Ma_M\,\,\,\,,
\,\,\,\,\forall x \in
\mathbb{R}^{2d}\,\,\,\mathrm{and}\,\,\,c^M \in \mathbb{Z}~. 
\end{equation}
One can choose a Cartesian reference frame and write the components of
each vector $a_M$ as $a^a_M$. Then the metric on $T^{2d}$
is\footnote{In our conventions the coordinates with the indices
  $M,N,\dots$ are parallel to the lattice defining the torus; they
  have period $2\pi \sqrt{\alpha'}$ and are referred to as the
  integral basis.} $G_{MN}= \sum_{a=1}^{2d}a^a_Ma^a_N$. The components
of each $a_M$ can be used to fill in the column vectors of a square
matrix $E^a_{\;M}\equiv a^a_M$; then, by construction, $E$ is a
vielbein matrix satisfying
\begin{equation}\label{vielbein}
G={}^tEE\,\,\,\,,\,\,\,\,1={}^tE^{-1}GE^{-1}~.
\end{equation}
The inverse matrix $E^{-1}$ instead has the dual vectors $\hat{a}^M$
as rows: $\sum_{a=1}^{2d}\hat{a}^M_{a} a^a_N=\delta^M_N$. Of course any
other matrix $E'=OE$ obtained by means of an orthogonal rotation $O$
of $E$ is a good vielbein matrix. 
We can also introduce complex coordinates and the complex vielbein
$\mathcal{E}$, such that
\begin{equation}\label{Eps}
\mathcal{E}=SE\,\,\,\,,\,\,\,\,S=\frac{1}{\sqrt{2}}\left(
\begin{array}{cc}
1 & \ii \\
1 & -\ii
\end{array}
\right) ~,
\end{equation}
where all the four blocks of $S$ are proportional to the $d \times d$
identity matrix. Two sets of complex coordinates are inequivalent if
they cannot be connected by a unitary transformation. Thus the
$SO(2d)$ ambiguity in the definition of the $E$'s implies that on the
same real torus there is a set of inequivalent complex structures
which is parametrized by $SO(2d)/U(d)$. Notice that in the complex
coordinates the flat metric $\mathcal{G}$ is off-diagonal
\begin{equation}\label{offmetric}
\mathcal{G}= {}^t \mathcal{E}^{-1}G\mathcal{E}^{-1}=\bar{S} S^{-1}= \left(
\begin{array}{cc}
0 & 1\\
1 & 0
\end{array} \right)~.
\end{equation}

In presence of a constant antisymmetric two-form, there is a
particular complex structure that plays a special role. Of course in
our applications this 2-form will be the gauge invariant combination
between the Kalb-Ramond $B$-field and $F$, the magnetic field on the
D-Branes: $\mathcal{F}=B+F$, thus we will indicate this antisymmetric
tensor with $\mathcal{F}$ already in this section (notice the
different convention with respect to \cite{Russo:2007tc}, as here the
factor of $2\pi \alpha'$ in front of $F$ is reabsorbed in the
definition of a dimensionless magnetic field).
Its expression in a real cartesian basis is
\begin{equation}\label{GF}
\mathcal{F}_c=\,{}^t\!E^{-1}\mathcal{F}E^{-1}=EG^{-1}\mathcal{F}E^{-1}~.
\end{equation}
and is related by a similarity transformation to the
combination $G^{-1}\mathcal{F}$ which will be the most relevant in the
following sections.
The antisymmetric matrix $ \mathcal{F}_c$  can be reduced to a
block-diagonal form
$\mathcal{F}_{\mathrm{b.d.}}$ by means of an orthogonal rotation $O_f$
(where the subscript is just to recall that this transformation
depends in general on $\mathcal{F}$)
\begin{equation}\label{Fbd}
\mathcal{F}_{\mathrm{b.d.}}=\left( 
\begin{array}{cc}
0 & \mathfrak{f}_d\\
-\mathfrak{f}_d & 0
\end{array}
\right)
=\,{}^t\!E_f^{-1}\mathcal{F}E_f^{-1}=E_fG^{-1}\mathcal{F}E_f^{-1}~,
\end{equation}
where $\mathfrak{f}_d$ is a $d \times d$ diagonal matrix with real
entries $\mathfrak{f}_{aa}$. The vielbein matrix $E_f=O_fE$ transforms
at the same time the metric $G$ into the identity and $\mathcal{F}$
into the block-diagonal matrix~\eq{Fbd}. Of course we can use the
vielbein matrix $E_f$ to introduce a particular set of complex
coordinates ($\mathcal{E}_f=SE_f$) which diagonalizes $G^{-1}\mathcal{F}$
\begin{equation}\label{F'}
\mathcal{F}^{(d)}=\mathcal{E}_fG^{-1}\mathcal{F}\mathcal{E}^{-1}_f=\left(
\begin{array}{cc}
-\ii \mathfrak{f}_d & 0\\
0 & \ii \mathfrak{f}_d
\end{array} \right)=\mathcal{G}\,{}^t\!
\mathcal{E}_f^{-1}\mathcal{F}\mathcal{E}_f^{-1}~,
\end{equation}
where ${}^t\!\mathcal{E}_f^{-1}\mathcal{F}\mathcal{E}_f^{-1}$ is a
block-diagonal matrix. From~\eq{F'} it is easy to see that, in this
complex basis, ${\cal F}$ is a $(1,1)$-form. In the following we will
always use, as Cartesian basis, the one defined by the vielbeins $E_f$
or ${\cal E}_f$, thus we will drop the subscripts without risk of
ambiguities.  Finally let us recall the definition of the complex
structure as the mixed tensor
\begin{equation}
\mathcal{I}=\ii dz \otimes \frac{\partial}{\partial z}-\ii
d\bar{z}\otimes \frac{\partial}{\partial \bar{z}}~.
\end{equation}
In the following sections we will need the expression of
$\mathcal{I}_a^{\phantom{a}b}$ in the integral basis, which is easily
derived by using the complex vielbein $\mathcal{E}$:
\begin{equation}\label{RealComplexStructure}
\mathcal{I}_M^{\phantom{M}N}=\left( {}^t\mathcal{E}
\right)_M^{\phantom{M}a}\mathcal{I}_a^{\phantom{a}b}\left(
  {}^t\mathcal{E}^{-1} \right)_b^{\phantom{b}N}~.
\end{equation}

\subsection{Closed strings on $T^{2d}$} \label{SectionVertex}

The coordinates of the closed strings propagating on $T^{2d}$ with a
constant background $B$-field can be written as a sum of left and right
handed fields, and the world-sheet is described by a free CFT. Then we
have the usual mode expansion
\begin{equation}\label{stringcoordinates}
  x^M_{\mathrm{cl}}(z,\bar{z}) =
  \frac{X^M_{\mathrm{cl}}(z)+\tilde{X}^M_{\mathrm{cl}}(\bar{z})}{2},  
  \,\,\,\,\mathrm{where}\,\,\,\,\,
X^M_{\mathrm{cl}}(z) = x^M-\ii \sqrt{2\alpha'}\alpha_0^M \ln
 z+\ii \sqrt{2\alpha'}\sum_{m \neq 0}\frac{\alpha_m^Mz^{-m}}{m}, 
\end{equation}
and the complexified world-sheet coordinates are
\begin{equation}\label{WScoordinates}
z=\ex{\tau+\ii \sigma}\,\,\,\,\,\,\mathrm{and}\,\,\,\,\,\,
\bar{z}=\ex{\tau-\ii \sigma}~,
\end{equation}
$\tau$ and $\sigma$ being the world-sheet time and spacial coordinates
respectively. Upon canonical quantization, the commutation relations
for the oscillators in Eq.~(\ref{stringcoordinates}) are
\begin{equation}\label{commutators}
\left[ \alpha_m^M,\alpha_{-n}^N \right]=n\delta_{m,n}G^{MN},
\,\,\,\,\,\, \mathrm{and}\,\,\,\,\,\, \left[ x^M,\alpha_0^N \right]=\ii
\sqrt{2\alpha'}G^{MN}~, 
\end{equation}
and similarly in the right moving sector.  The allowed winding and
Kaluza-Klein modes are encoded in the Narain lattice which depends on
the background fields $G$ and $B$ as follows
\begin{equation}\label{zeromodes}
\alpha_0^M=\frac{G^{MN}}{\sqrt{2}} \left[ \hat{n}_N+ \left(
    G_{NN'}-B_{NN'} \right) \hat{m}^{N'}
\right]\,\,\,\,,\,\,\,\,\tilde{\alpha_0}^M=\frac{G^{MN}}{\sqrt{2}}
\left[ \hat{n}_N- \left( G_{NN'}+B_{NN'} \right) \hat{m}^{N'} \right],
\end{equation}
where $\hat{n}_M$ and $\hat{m}^M$ are respectively the Kaluza-Klein
and winding numbers operators.

Let us now consider the interacting theory and pay particular
attention to the phases necessary when the target space is compact.
The basic building block is the tree-level coupling among three
generic closed strings (Reggeon vertex). Since there is no preferred
ordering of three points on a sphere, this vertex must be invariant
under the action of the permutation group exchanging any of the
punctures. The usual emission vertex valid in the uncompact space does
not have this property when it is naively generalized to the $T^{2d}$
case. This is a well known issue, related to the compactness of the
target space, and it is a consequence of the logarithmic branch cut of
the bosonic Green function. Similar problems arise also in the usual
formalism of the vertex operators describing the emission of
particular on-shell string states (see, for instance, Sect.~8.2
in~\cite{Polchinski:1998rq}). In order to eliminate this branch cut
one has to add suitable cocycle factors to the usual expression of the
vertex operators. Here we tackle this issue exactly in the same way by
generalising these cocycle factors to the Reggeon vertex formalism.

In order to see that the usual $3$-string vertex for closed strings is
not invariant under the permutation of the external states when the
target space is a torus, it is sufficient to focus on the zero-modes
contribution. The explicit form of the full vertex $V_3$ can be found,
for instance, in \cite{Russo:2007tc} and its zero-mode part is simply
\begin{equation}\label{ExpLn}
  \exp \left[\sum_{j>i=0}^2 \a^i_0 \ln(z_j-z_i)G \a^j_0 +
    \sum_{j>i=0}^2 \tilde\a^i_0 \ln(\bar{z}_j-\bar{z}_i)G \tilde\a^j_0 
  \right], 
\end{equation}
where the $z_i~(i=0,1,2)$ are the positions of the three punctures on
the sphere (which is represented as the compactified complex plane);
the upper index identifies one of the three external states, and all
space-times indices have been suppressed. The oscillator part of the
$3$-string vertex is invariant under the exchange of the strings $1
\leftrightarrow 2$, while the zero-mode contribution~(\ref{ExpLn})
gets a phase given by
\begin{equation}\label{phase}
\exp \left[ -\ii\pi \left( \a_0^1G\a_0^2-\tilde{\a}_0^1G\tilde{\a}_0^2
  \right) \right] = \exp \left[ -\ii \pi \bigg(
  \hat{n}_1\hat{m}_2+\hat{m}_1 \hat{n}_2 \bigg) \right], 
\end{equation}
where we have used~\eq{zeromodes}. By using this result, we can build a new
invariant vertex with a cocycle factor that compensates for the
phase~(\ref{phase}). A possible choice for this cocyle factor is
\begin{equation}\label{vertex}
V^c_3=V_3 \exp \left[ \frac{\ii \pi}{2} \bigg(
  \hat{n}_1 \hat{m}_2 -\hat{m}_1 \hat{n}_2 \bigg) \right]~. 
\end{equation}
By using the conservation of the Kaluza-Klein and winding numbers of
the emitted strings, the vertex~(\ref{vertex}) is now easily shown to
be invariant under the full permutation group acting on the three
punctures.

However, the cocycle factor added seems to break the vertex invariance
under T-Duality transformations. For instance, there is a particular
T-duality transformation that exchanges the Kaluza-Klein and winding
operators. If its effect could be simply written as $\hat{n}
\leftrightarrow \hat{m}$, as it is usually done, then we could have
that $V_3^c \to -V_3^c$ for certain external states. In order to
clarify this point, let us see the transformation properties of
$V_3^c$ under a generic T-duality transformation. These
transformations can be encoded in a $O(2d,2d,\mathbb{Z})$
matrix~\cite{Giveon:1994fu}
\begin{equation}\label{Tduality}
T= \left(
\begin{array}{cc}
a & b\\
c & d
\end{array}
\right)\,\,\,\,,\,\,\,\,\,\, \mathrm{with} \,\,\,\,\,\,
a\, {}^t\! b+ b\, {}^t\! a=0
\,\,\,,\,\,\,
c \,{}^t\! d+d\, {}^t\! c=0\,\,\,,\,\,\,
a\, {}^t\!d+b\,{}^t\!c=1~,  
\end{equation}
where $a$, $b$, $c$ and $d$ are $2d \times 2d$ matrices and their
constraints follow from the group definition, namely  
\begin{equation}\label{symplectic}
J=TJ\, {}^t T\,\,\,\,\,\,,\,\,\,\,\,\,
J=\left(
\begin{array}{cc}
0 & 1\\
1 & 0
\end{array}
\right).
\end{equation}
Let us observe that, as $J=J^{-1}$, by inverting the relation above,
one finds that also ${}^tT$ is an $O(2d,2d,\mathbb{Z})$ matrix with
similar constraints imposed on its entries. The duality acts as
follows on the geometrical background
\begin{equation}\label{E'}
G'+B'=\left[ a(G+B)+b \right] \left[ c(G+B)+d \right] ^{-1}
\end{equation}
and on the string oscillators
\begin{equation}\label{a'}
\alpha_n =T_+\alpha'_n ~~,~~~~
\tilde{\alpha}_n  = T_-\tilde{\alpha'}_n~,
\end{equation}
where
\begin{eqnarray}\label{a''}
T_{\pm} =
\left[c(\pm G+  B)+ d \right] ^{-1}. 
\end{eqnarray}
Notice that one can prove \cite{Giveon:1994fu} that
${}^tT_{\pm}GT_{\pm}=G'$. Focusing now on the zero-modes, we can
derive from~(\ref{a'}) the action of the T-duality transformation on
the winding and Kaluza-Klein operators
\begin{equation}\label{n'm'}
\hat{n}={}^td\hat{n}'+{}^tb\hat{m}'\,\,\, , 
\,\,\,\,\hat{m}={}^tc\hat{n}'+{}^ta\hat{m}'
~~~\Longleftrightarrow~~~
\hat{n}'=a\hat{n}+b\hat{m} \,\,\, , 
\,\,\,\, \hat{m}'=c\hat{n}+d\hat{m}
\end{equation}
Then the transformation law for the vertex~(\ref{vertex}) is
\begin{eqnarray}\label{V'}
  && V^c_3 = V^{c'}_3\ex{\ii \pi \left[
      {}^t\hat{n}_1'd^tc\hat{n}_2' + {}^t\hat{m}_1'b^ta\hat{m}_2' + 
 {}^t\hat{n}_1'd^ta\hat{m}_2' +
   {}^t\hat{m}_1'b^tc\hat{n}_2' - {}^t\hat{n}_1'\hat{m}_2' \right]}=\\ 
 \nonumber && = V^{c'}_3 \prod_{j=0}^2 \bigg\{ \ex{ \ii \pi
 \left[ \sum_{M<N}\left( (\hat{n}'_j)_M (d {}^tc)^{MN} (\hat{n}'_j)_N +
(\hat{m}'_j)^M (b {}^ta)_{MN} (\hat{m}'_j)^N \right) +
{}^t\hat{m}'_jb {}^tc \hat{n}'_j \right]} \bigg\}, 
\end{eqnarray} 
where in the last line we used of the conservation of windings and
momenta. Thus the vertex~(\ref{vertex}) is fully symmetric under
permutations of the external states, but is not invariant under the
T-Duality transformations~\eq{a'}. However, it is interesting to
notice that the phase (which is actually just a sign) generated by the
transformations~\eq{a'} can always be written as a product of three
terms each one depending only a single external state, as it is done
in the second line of~\eq{V'}.  This means that the invariance of the
vertex~(\ref{vertex}) under T-Duality can be restored, provided that
we introduce the appropriate cocycle also in the T-duality
transformations, as a generalisation of the standard rules discussed
above and in~\cite{Giveon:1994fu}. In fact it is sufficient to
postulate that the closed string states transform according
to~\eq{n'm'} {\em and} also acquire the same phase in the curly
brackets of (\ref{V'})
\begin{equation}
  \label{ts}
  \ket{n,m} \to
 \ex{ \ii \pi
 \left[ \sum_{M<N}\left( (\hat{n}')_M (d {}^tc)^{MN} (\hat{n})_N +
(\hat{m}')^M (b {}^ta)_{MN} (\hat{m}')^N \right) +
{}^t\hat{m}'b {}^tc \hat{n}'\right]} 
\; \ket{n',m'}~.
\end{equation}
We will see in the following section that this is indeed the case when
considering boundary states describing magnetised D-Branes as dual for
instance to configurations of purely Dirichlet or intersecting
D-Branes.

It is not difficult to generalize the analysis above to the case of a
Reggeon vertex describing the interaction of many closed strings. This
can be obtained just by gluing together the 3-point
vertices~\eq{vertex} and the result is
\begin{equation}\label{Nvertex}
V^c_{N+1}=V_{N+1} \exp \left[ \frac{\pi \ii}{2} \sum_{j>i=1}^{N}
  \bigg(\hat{n}_i \hat{m}_j -\hat{m}_i \hat{n}_j 
  \bigg) \right]~.
\end{equation}
Here $V_{N+1}$ is the standard vertex, where the cocycles are ignored
and, as before, the indices $i,\;j=0,1,\ldots,N$ label the $N+1$
external states.

\subsection{The Boundary State for a wrapped magnetized D-Brane}\label{SectionBoundaryState}

In this section we study, from the closed string point of view, the
space-filling magnetised D-Branes with generic wrapping numbers on the
torus cycles (see~\cite{Pesando:2005df,DiVecchia:2006gg} for previous
works on this subject). In the closed string sector D-Branes are
described by boundary states $\ket{B_F}$ that enforce an
identification between the left and right moving modes (for a review
of the boundary state formalism in the Ramond Neveu-Schwarz formalism
see~\cite{DiVecchia:1999rh,DiVecchia:1999fx}). For magnetized D-branes
we have
\begin{equation}\label{Identification}
\left[(G+\mathcal{F}) \alpha_n +
  (G-\mathcal{F}) \tilde{\alpha}_{-n} \right]
\ket{B_F}=0,\,\,\,\,\,\,\forall n \in \mathbb{Z}~,
\end{equation}
where we used the gauge invariant combination of the Kalb-Ramond
$B$-field and the field strength $F$ on the D-brane:
$\mathcal{F}=B+F$. In the integral basis, the magnetic field is
quantized as a consequence of the compactness of the torus,
\begin{equation}\label{Fquantization}
F_{MN}=\frac{p_{MN}}{w_Mw_N}~,
\end{equation}  
$p_{MN}$ being an integer matrix and $w_M$ being the wrapping numbers
of the D-Brane along the cycles of the torus. Clearly
Eq.~\eq{Identification}, being a set of linear constraints, fixes the
form of the boundary state up to an overall factor that can also
depend on the Kaluza-Klein and winding operators. We will now show
that the boundary state for a magnetised D-Brane does indeed contain
non-trivial phases that depend on the winding numbers and on the field
$F$; moreover it turns out that this phase is strictly related to the
phase that closed string states acquire under T-Duality.
Following~\cite{Callan:1988wz}, we consider the gauge field
contribution to the action in the string path-integral as an
interaction term that acts on the standard boundary state for an
unmagnetised D-Brane wrapping the $T^{2d}$. We want to derive the
dependence of $\ket{B_{F}}$ on the magnetic field by applying the
usual path ordered Wilson loop operator
$\mathrm{P}[\exp{(\frac{\ii}{2\pi\alpha'}\oint A dx)}]$ to
$\ket{B_{F=0}}$. The computation was performed in~\cite{Callan:1988wz}
in a Minkowski flat target space, while the novelty of the present
calculation\footnote{An analogous calculation is performed
  in~\cite{Igor:2008}; we thank I. Pesando for letting us know 
  their results before publication.}  is the compactness of the torus
wrapped by the D-Branes.  As the non-zero modes contribution to the
boundary state is not affected by the shape of the target space, we
will mostly focus on the zero-modes. Of course, our aim is to
determine all the $F$-dependent terms that cannot be fixed
from~\eq{Identification}. In order to do so we have to pay some care
to the definition of the Wilson loop operator.

\subsubsection{Gauge bundles and gauge invariant Wilson loop}

Any gauge potential for~\eq{Fquantization} will involve a linear, and
thus non-periodic, function.  Let us start from the simpler case
$w_M=1$, $\forall M$. As usual, we need to compensate for the
non-periodicity of $A$ by introducing a set of gauge transformations
$U_N$. Each $U_N$ encodes the gluing conditions for the gauge
potential between two copies of the torus that are adjacent along the
$N$-th direction in the covering space. So the gauge potential living
on a D-Branes world volume wrapping a cycle of the compactification
torus must satisfy 
\begin{equation}\label{APeriodicity}
A_M\left( x+2\pi \sqrt{\alpha'}a_N \right)=U_N(x)\left( 2\pi
  \alpha' \ii \partial_M +  A_M(x) \right) U_N^{\dagger}(x)~, 
\end{equation}
where $a_N$ denotes the $N$-th cycle of the torus. In the case under
analysis, all gauge transformations $U_N$ belong to $U(1)$ and so the
formula above can be further simplified. We choose not to do it, so as
to keep the equations~\eq{APeriodicity}--\eq{U'} valid also for the
non-abelian generalization that we will need once we reintroduce
multiple wrappings. The background gauge field (gauge bundle) is
properly defined by Eq.~\eq{Fquantization} together with the set of
$U_N$'s. In order to have a consistent bundle, the gluing matrices
must satisfy the overlap condition
\begin{equation}\label{consistency}
U_N^{\dagger}(x)U_M^{\dagger}\left( x+2\pi \sqrt{\alpha'}a_N \right) 
U_N \left( x+2\pi \sqrt{\alpha'}a_M \right)U_M(x)=1~.
\end{equation}
All fields charged under the gauge potential have to obey periodicity
conditions similar to~\eq{APeriodicity}. For instance, fields
transforming in the fundamental ($\Phi$) or in the adjoint ($\Psi$)
representation must satisfy
\begin{equation}\label{FunPeriodicity}
\Phi\left( x+ 2\pi\sqrt{\alpha'}a_N \right)=U_N(x)\Phi(x)~~,~~~~
\Psi\left( x+2\pi \sqrt{\alpha'}a_N \right)=U_N(x)\Psi(x)U_N^{\dagger}(x)~.
\end{equation}
As a consequence of Eq.~(\ref{FunPeriodicity}), under a generic gauge
transformation $\gamma(x)$, the gluing matrices $U_N$ transform in
the following fashion
\begin{equation}\label{U'}
U_N(x) \rightarrow \gamma \left( x+ 2\pi \sqrt{\alpha'}a_N  \right) 
U_N(x) \gamma^{\dagger}(x)~.
\end{equation}
Notice that there are no restrictions on $\gamma(x)$ and in particular
it does not have to be periodic. So the form of the $U_N$'s can change
under a gauge transformation.

We now focus on the definition of the Wilson loop operator we need
for the computation of the magnetized boundary state. Let us consider
a path $c$ connecting two points that are separated by the lattice
vector $\sum_{L=1}^{2d}m_La_L$, $m_L \in \mathbb{Z}$ (which means that
they are identified on the torus). In order to be consistent with our
convention for $\mathcal{F}$, this path will start from
$x+2\pi\sqrt{\alpha'}\sum_{L=1}^{2d}m_La_L$ and end in $x$. Then it is
clear that the naive path ordered Wilson loop operator is not gauge
invariant\footnote{and even depends on the initial point $x$ of the
  path $c$.}, but 
transforms as
\begin{equation}\label{Wilson'}
\mathrm{P}[\ex{\ii \int_c A dx}] \rightarrow \gamma(x)\;
\mathrm{P}[\ex{\ii \int_c A dx}] \; \gamma^{\dagger}\left( x+
  2\pi\sqrt{\alpha'}\sum_{L=1}^{2d} m_L a_L \right) ~.
\end{equation}
We can recover a gauge invariant object if we multiply the Wilson
loop~\eq{Wilson'} by a sequence of $U$'s which forms a discretized
version of the path $c$. By using~\eq{consistency} we can choose to
collect together all shifts along the direction $K=1$, then those
along $K=2$ and so on. In formulae we have
\begin{equation}\label{InvariantWilson}
\left[ \prod_{K=1}^{2d}\prod_{m=0}^{m_K-1}U_K \left(
    x+2\pi \sqrt{\alpha'}\sum_{L=1}^{K-1}m_La_L+2\pi
    \sqrt{\alpha'}ma_K \right)\right]
   \mathrm{P}[\ex{\frac{\ii}{2\pi\alpha'} \int_c A dx}]~, 
\end{equation}
where only the values $m_K\geq 1$ are relevant (if we have $m_K=0$ for
certain $K$, then the corresponding $U_K$ does not appear in the
product). By using~\eq{U'} and~\eq{Wilson'} (and~\eq{APeriodicity}), 
one can check that~\eq{InvariantWilson} is invariant under an
arbitrary $U(1)$ gauge transformation $\gamma(x)$ (and does not depend
on the initial point $x$ of the path $c$).

\subsubsection{Wrapped D-Branes as non abelian gauge bundles}

A D-brane with multiple wrappings ($w_M >1$ for some $M$) is better
described~\cite{Guralnik:1997sy,Guralnik:1997th} in terms of a
non-trivial gauge bundle on the torus $T^{2d}$~\cite{tHooft:1981sz}.
In the D-brane language this amounts to considering a set of
$w=\prod_{M=1}^{2d}w_M$ coincident D-branes with the same gauge field
strength~\eq{Fquantization}, but with non-trivial transition matrices
$U_M$. The non-abelian character of the configuration is encoded in
the $U_M$'s that, as consequence of~\eq{FunPeriodicity}, glue together
the various D-branes in a single wrapped object. In absence of
magnetic fields, this can be easily seen by choosing as $U_M$ the
following $w_M \times w_M$ transition matrix
\begin{equation}\label{P}
U_M=P_{w_M \times w_M}=\left(
\begin{array}{cccccc}
0 & 1 & 0 & 0 & \cdots & 0\\
0 & 0 & 1 & 0 & \cdots & 0\\
0 & 0 & 0 & 1 & \cdots & 0\\
\vdots & \vdots & \vdots & \vdots & \ddots & \vdots \\
1 & 0 & 0 & 0 & \cdots & 0
\end{array} \right)~.
\end{equation}
Notice that, when $F=0$, a D-Brane wrapped $w_M$ times along the
$M$-th cycle of the torus can be smoothly deformed, at the classical
level, into $w_M$ coincident
branes~\cite{Polchinski:1996fm,Hashimoto:1996pd}. This means that
there is a family of flat gauge bundles interpolating between~\eq{P}
and $U_M=1$.

In order to introduce the effect of the magnetic field
(\ref{Fquantization}) on the D-Brane, it turns out to be convenient to
choose a fundamental cell of the torus lattice where this field is in
a block-diagonal form. This can be always done~\cite{Griffiths} (see
also the Appendix for a more pedestrian proof), so from now on we take
\begin{equation}\label{Fblock}
F=\left(
\begin{array}{ccccc}
0 & \frac{p_1}{W_1} & 0 & 0 & \cdots\\
-\frac{p_1}{W_1} & 0 & 0 & 0 & \cdots\\
0 & 0 & 0 & \frac{p_2}{W_2} & \cdots\\
0 & 0 & -\frac{p_2}{W_2} & 0 & \cdots\\
\vdots & \vdots & \vdots & \vdots & \ddots
\end{array} \right)~,
\end{equation}
where $p_{\alpha} \in \mathbb{Z}$ and $W_{\alpha} \in
\mathbb{Z}-\{0\}$, $\forall \alpha=1,\dots,d$. Notice that even if the
field $F$ now describes a direct product of $d$ $T^2$'s inside the
$T^{2d}$, the compactification is in general non factorizable as a
consequence of the form of the metric.  The $p_{\alpha}$'s can be
interpreted as the Chern classes of the magnetic fields while the
$W_{\alpha}$'s are the products of the couples of wrapping numbers on
each of the $T^2$'s inside $T^{2d}$.

Two comments are in order now. First, if $p_\alpha$ and $W_\alpha$ are
not co-prime, the configuration can again be smoothly deformed, at the
classical level, to a new configuration with co-prime $p'_\alpha$ and
$W'_\alpha$ and $p'_\alpha/W'_\alpha=p_\alpha/W_\alpha$. Second it is
not necessary to specify all the wrappings of the brane along each
cycle of the torus. In fact it is possible to show that configurations
of magnetized D-Branes with the same product of wrappings along the
pairs of cycles of the $T^2$'s inside the $T^{2d}$ defined by the form
of the field (\ref{Fblock}) are equivalent. Indeed the transition
matrices defining the gauge bundle of a brane wrapped along a $T^2$
are related by a gauge transformation if they describe branes with the
same Chern class $p$ and the same product of the wrappings $W$. Thus
we can choose to wrap the branes $W_{\alpha}$ times along the even
directions $x_M\equiv x_{2\alpha}$ only. We will also make the
following gauge choice for the gauge potential~(\ref{Fblock})
\begin{equation}\label{gauge}
A_{M \equiv 2\alpha}(x)=\frac{p_{\alpha}}{W_{\alpha}}x_{2\alpha-1}+
2\pi\sqrt{\alpha'}C_{2\alpha-1}
\,\,\,\,\,\,\mathrm{and}\,\,\,\,\,\,   
A_{M \equiv 2\alpha-1}(x)=2\pi\sqrt{\alpha'}C_{2\alpha}~,\,\,\,\,\,\, 
\forall \alpha=1,\dots d\,,
\end{equation}
where we have introduced non-zero Wilson lines $2\pi
\sqrt{\alpha'}C_M$, with $C_M$ adimensional. In this way the non
abelian gauge bundle describing the magnetized wrapped D-Brane is
characterized by
\begin{eqnarray}\label{U}
U_{2\alpha}(x) &=& 1_{W_1 \times W_1} \otimes \ldots \otimes
P_{W_{\alpha} \times W_{\alpha}} \otimes 1_{W_{\alpha+1} \times
  W_{\alpha+1}} \otimes \ldots \otimes 1_{W_{d} \times  W_{d}}
 \\ \nonumber 
U_{2\alpha-1}(x) &=& 1_{W_1 \times W_1} \otimes \ldots \otimes
(Q_{W_{\alpha}  \times W_{\alpha}})^{p_{\alpha}} \otimes 1_{W_{\alpha+1} \times
  W_{\alpha+1}} \otimes \ldots \otimes 1_{W_{d} \times  W_{d}}
~\ex{^{\frac{\ii}{\sqrt{\alpha'}}\frac{p_{\alpha}}{W_{\alpha}}x_{2\alpha}}}\,,
\end{eqnarray}
where $Q_{W_{\alpha} \times W_{\alpha}}=\mathrm{diag}\left\{ 1,\ex{
    2\pi\ii /W_{\alpha}},\dots,\ex{2\pi \ii (W_{\alpha}-1)/W_{\alpha}}
\right\}$.  Notice that the form of the transition matrices $U_M$ is
not affected by the presence of the Wilson-lines as for them the
relation (\ref{APeriodicity}) is trivially satisfied by the identity
gluing matrix. 

The generalization of~\eq{InvariantWilson} to this non-abelian setup
is straightforward and the operator we need to use to derive the
magnetized boundary state from the unmagnetized one reads
\begin{equation}\label{WilsonOperator}
{\cal O}_A = \Tr \left\{
\left[ \prod_{K=1}^{2d}\prod_{m=0}^{\hat{m}_K-1}U_K
  \left(x_{\rm cl}+\sum_{L=1}^{K-1}2\pi
    \sqrt{\alpha'}\hat{m}_La_L+2\pi \sqrt{\alpha'}m a_K \right)
\right] \mathrm{P}[\ex{\frac{\ii}{2\pi\alpha'} \int_c A dx_{\rm
    cl}}] \right\}\,,  
\end{equation}
where the $x_{\rm cl}^M$'s are the usual string
coordinates~\eq{stringcoordinates} and the $\hat{m}_M$'s are the
operators that read the winding numbers of the closed string states.
In this non-abelian generalization we have to put $U(x)$ at the right
hand of the sequence and then follow the order determined by the path
$c$.

\subsubsection{Computation of the Boundary State}

We can now compute the action of ${\cal O}_{A}$ on the unmagnetized
boundary state $\ket{B_{F,C}} = {\cal O}_A \ket{B_{A=0}}$. The
unmagnetized boundary state for a wrapped D-Brane is found from the
one of an unwrapped brane (see~\cite{DiVecchia:1999fx} and references
therein) by applying the same operator (\ref{WilsonOperator}) with the
choice
\begin{equation}
A_M=0\,\,\,\,\,\, \mathrm{and}\,\,\,\,\,\, 
U_K=1_{w_1 \times w_1} \otimes \ldots \otimes P_{w_K \times
  w_K} \otimes 1_{w_{K+1} \times
  w_{K+1}} \otimes \ldots \times 1_{w_{2d} \times w_{2d}}~,
\end{equation}
with $P$ defined as in Eq.~(\ref{P}). In this case the trace
in~\eq{WilsonOperator} reads
\begin{equation}
\mathrm{Tr}\left[ \prod_{K=1}^{2d}U_K^{\hat{m}_K} \right]
\end{equation}
and it is different from zero only when the windings of the emitted
closed strings are integer multiples of the wrappings of the D-Brane
on each cycle of the torus. Hence only these states couple to the
wrapped D-brane, as expected, and we have
\begin{equation}\label{BF0}
\ket{B_{A=0}}=\sqrt{\mathrm{Det}(G+B)}\sum_{m^M \in Z} \prod_{n=1}^{\infty}
\ex{-\frac{1}{n}\tilde{\a}_n^{\dagger}G R_0\,\a_n^{\dagger}}
\ket{0;w_M m^M}~,
\end{equation}
where there is no sum understood over the repeated index $M$;
$R_0=(G-B)^{-1}(G+B)$ is the identification matrix between left and
right moving oscillators and depends on the geometric background of
the torus; finally the ket $\ket{0;w_Mm^M}$ represents the closed
string vacuum state with zero Kaluza-Klein momenta and winding numbers
equal to $w_M m^M$ for $M=1,2,...,2d$.

It is easy to begin by turning on only the Wilson lines and to keep
vanishing magnetic fields on the D-brane world volume. We have just to
isolate the Wilson line contribution to $\mathcal{O}_A$
in~\eq{WilsonOperator} when acting on $\ket{B_{A=0}}$, namely
\begin{equation}\label{WL}
\ket{B_{C}}=
\ex{\frac{\ii}{\sqrt{\alpha'}} \int_c C \cdot dx}~
\ket{B_{A=0}}=\ex{2\pi \ii C \cdot \hat{m}}\ket{B_{A=0}}~.
\end{equation}   
The explicit evaluation of the contributions of the magnetic fields
$F$ is longer. It can be split into the zero and non-zero mode part
of the string coordinate $x_{\rm cl}(z,\bar{z})$ defined
in~\eq{stringcoordinates}.  Let us focus on the zero-mode part of the
computation, since the non-zero mode contribution has just the effect
of replacing $B$ with $\mathcal{F}$ in~(\ref{BF0}),
see~\cite{Callan:1988wz}. A first result is that the magnetized
boundary state couples only with the closed strings whose windings in
the $\alpha$-th $T^2$ (as defined by the form of the magnetic field
in~(\ref{Fblock})) are integer multiples of $W_{\alpha}$. This is
again a consequence of the trace in~\eq{WilsonOperator} where the
transition matrices are defined as in Eq.~(\ref{U}). It is convenient
to define the $2d \times 2d$ matrix
\begin{equation}\label{wma}
w=\left(
\begin{array}{ccc}
W_1 & 0 & \cdots\\
0 & W_2 & \cdots\\
\vdots & \vdots & \ddots
\end{array} \right)~\otimes 1_{2\times 2}
\end{equation}
and use $w m$ to indicate the $2d$ column vector containing the
windings of the closed strings emitted by the magnetized D-Brane. We
can also see how the action of ${\cal O}_A$ yields the relation
between windings and Kaluza-Klein numbers
\begin{equation}\label{nFm}
\hat{n} = -F \hat{m}~,
\end{equation}
which is usually derived from the identification imposed by
Eq.~(\ref{Identification}) on the closed string zero-modes. For this,
it is sufficient to focus on the zero-modes contribution linear both
in the position operator $\hat{q}^M=\left(x^M+\tilde{x}^M \right)/2$
and in the oscillators $\a_0$ or $\tilde{\a_0}$. Using the form of $F$
in Eq.~(\ref{Fblock}) with the gauge choice~(\ref{gauge}) and the
transition matrices~\eq{U}, we can evaluate this contribution as
follows
\begin{eqnarray}
\nonumber \ket{B_F} & \sim &
\ex{\frac{\ii}{\sqrt{\alpha'}}\sum_{\alpha=1}^d
 \frac{p_{\alpha}}{W_{\alpha}}\hat{m}_{2\alpha-1}\hat{q}_{2\alpha}}
 \ex{-\ii \frac{\sqrt{\alpha'}}{2\sqrt{2}}
  \sum_{\alpha=1}^d\int_{0}^{2\pi}d\sigma \frac{1}{2\pi
  \alpha'}\frac{p_{\alpha}}{W_{\alpha}}(x_{2\alpha-1}+\tilde{x}_{2\alpha-1})
  (\a_0^{2\alpha}-\tilde{\a}_0^{2\alpha})}\ket{0;wm}=\\ 
 & = & 
 \ex{\ii
  \sum_{\alpha=1}^d\frac{p_{\alpha}}{W_{\alpha}}\hat{m}_{2\alpha-1}
 \frac{\hat{q}_{2\alpha}}{\sqrt{\alpha'}}}\ex{-\ii \sum_{\alpha=1}^n
 \frac{p_{\alpha}}{W_{\alpha}}\frac{\hat{q}_{2\alpha-1}}{\sqrt{\alpha'}}
 \hat{m}_{2\alpha}}\ket{0;wm}=
 \ket{-Fwm,wm}~.
\end{eqnarray} 
Notice that at this stage we can forget about the path-ordering in the
Wilson operator $\mathcal{O}_A$ and explicitly perform the integration
in the first line of the previous equation, as the zero-mode
contributions of the string fields entering the Wilson loop in
$\mathcal{O}_A$ commute with each other at different values of
$\sigma$. Finally let us consider the terms quadratic in the
zero-modes $\alpha_0$ and $\tilde{\alpha}_0$ that follow from the
standard Wilson loop exponential in Eq.~(\ref{WilsonOperator}),
\begin{eqnarray} \label{QuadraticZeroModes}
&&
\ex{-\frac{\ii}{2}\sum_{\alpha=1}^d\int_{0}^{2\pi}d\sigma
\frac{1}{2\pi}\frac{p_{\alpha}}{W_{\alpha}}\left(
  \a_0^{2\alpha-1}\a_0^{2\alpha}-\tilde{\a}_0^{2\alpha-1}
  \a_0^{2\alpha}-\a_{0}^{2\alpha-1}\tilde{\a}_0^{2\alpha}+
  \tilde{\a}_0^{2\alpha-1}\tilde{\a}_0^{2\mu}\right) \sigma}
   \ket{-Fwm;wm}=
 \\ \nonumber && 
= \ex{-\ii \pi \sum_{\alpha=1}^d W_{\alpha}p_{\alpha}
 m_{2\alpha-1}m_{2\alpha}}\ket{-Fwm;wm} =
\ex{-\ii\pi\sum_{M<N}\hat{m}^M F_{MN} \hat{m}^N}
 \ket{-Fwm;wm}~.
\end{eqnarray}
The phase in~(\ref{QuadraticZeroModes}) could not have been deduced
just by looking at the constraints~(\ref{Identification}). So the
expression for the boundary state describing a magnetised D-Brane is
\begin{eqnarray}\label{WLBF}
\ket{B_{F,C}}=  \sqrt{\mathrm{Det}(G+\mathcal{F})} 
  & \sum_{m \in \mathbb{Z}^{2d}} &\!
  \ex{-\ii\pi\!\sum_{M<N}\hat{m}^MF_{MN}\hat{m}^N}
  \, \ex{2\pi \ii C \hat{m}}
 \\ \nonumber & \times & \!\!\!
   \left[\prod_{n=1}^{\infty}\ex{-\frac{1}{n}\tilde{\a}_n^{\dagger}
   G R\, \a_n^{\dagger}} \right]
 \ket{-Fwm ,wm}~,  
\end{eqnarray} 
where the identification matrix $R$ is
\begin{equation}
  \label{eq:imr}
 R=(G-\mathcal{F})^{-1}(G+\mathcal{F})~.  
\end{equation}
Even if the phase in Eq.~\eq{WLBF} has been calculated for a block
diagonal $F$, in the Appendix we will prove that Eq.~\eq{WLBF} holds
for a generic $F$ (see Eq.~\eq{Fquantization}), a part for  possible
half integer shifts in the Wilson line.

Let us now analyze how Eq.~\eq{WLBF} transforms under an
$O(2d,2d,\mathbb{Z})$ transformation~(\ref{Tduality}). Generically,
after the T-duality, we have a new magnetized D-brane with
\begin{equation}\label{Fduality}
F' = (aF - b)(-cF +d)^{-1}~,~~~~R'=T_-^{-1}R T_+,
\end{equation}
as it can be seen by using the relations in~(\ref{n'm'}); moreover,
since $F$ is an antisymmetric matrix, also $F'$ is antisymmetric,
thanks to constraints in~(\ref{Tduality}). If the combination
$(-cF+d)$ in~\eq{Fduality} is not invertible, then the transformed
D-brane will have some direction with Dirichlet boundary conditions.
For instance, we can check that any magnetized brane can be easily
related to a lower dimensional D-Brane at angles via T-Duality. First
let us put the magnetic field in the block-diagonal form, as in
Eq.~(\ref{Fblock}). Then, we T-dualize the even direction of each of
the $T^2$'s defined by $F$ inside the $T^{2d}$, that is, we choose the
following $O(2d,2d,\mathbb{Z})$ matrix
\begin{equation}
a=d= 1_{d \times d} \otimes \left(
\begin{array}{cc}
1 & 0\\
0 & 0
\end{array} \right)\,\,\,\,\,\, 
\mathrm{and}\,\,\,\,\,\, 
b=c= 1_{d \times d} \otimes \left( 
\begin{array}{cc}
0 & 0\\
0 & 1
\end{array} \right)~.
\end{equation}
By using the action of the duality transformations of the Kaluza-Klein
and winding operators, as in Eq.~(\ref{n'm'}), the phase
(\ref{QuadraticZeroModes}) can be rewritten as
\begin{equation}
  \ex{-\ii \pi
    \sum_{\alpha=1}^d\frac{p_{\alpha}}{W_{\alpha}}
  (W_{\alpha}^2)m_{2\alpha-1}m_{2\alpha}}
  = \ex{-\ii \pi \sum_{\alpha=1}^d\hat{m}'_{2\alpha}\hat{n}'_{2\alpha}}~.
\end{equation}
Notice that this phase exactly compensates for the one that the closed
string states in the ket of~\eq{WLBF} acquire under T-Duality, as
stated in Eq.~(\ref{ts}). Thus we see that the boundary states for
purely geometrical configurations of D-Branes (like brane intersecting
at angles) do not contain any non trivial phase depending on the
emitted closed strings zero-modes, as expected. By reinstating the
$g_s$ dependence and using the results of~\cite{Myers:1999ps}, it is
possible to show that also the prefactor
$\sqrt{\mathrm{Det}(G+\mathcal{F})}$ transforms into the one expected
for the boundary state of a D-Brane at angle.

It is also possible to transform any magnetized D-brane into a D-brane
with Dirichlet boundary conditions along all the coordinates of the
torus. In this case, the matrices $c$ and $d$ defining the T-duality
are related to the magnetic field $F= c^{-1}d$. When $F$ is
block-diagonal~(\ref{Fblock}),
one can choose $c=w$ given by ~(\ref{wma}) and easily build integer matrices
$a$ and $b$ satisfying Eqs.~(\ref{Tduality}) using the fact that 
the wrappings  $W_\alpha$  and the Chern numbers 
$p_\alpha$ are coprime. Exactly as in the previous example, the phase of the
magnetized boundary state cancels against the phase generated by the
T-duality transformation~(\ref{ts}). Thus one recovers the standard
form of a Dirichlet boundary state, where the identification matrix is
simply $R=-1$.

Finally the transformations on the closed strings
zero-modes~(\ref{ts}) show explicitly that the Wilson lines
in~(\ref{WLBF}) are related to the positions of the D-Brane if the
dualized directions have Dirichlet boundary condition. For the case of
the D-branes at angles, in each of the two-dimensional tori inside the
$T^{2d}$, half of the components of the Wilson lines becomes positions
and the remaining ones are still interpreted as residual Wilson lines
on the dualized brane. When the D-brane is transformed into a point in
the compact space, then all components of the Wilson lines are
geometrized into positions of the dual D-brane.

\sect{The twisted partition function}\label{SectionAmplitude}

In this section we will compute the bosonic contribution to the open
string twisted partition function for multiply wrapped D-branes.  The
planar partition function $Z_g(F)$ is obtained by starting from the
tree-level vertex~(\ref{Nvertex}) and sewing the external legs with
boundary states describing magnetized D-Branes with Wilson lines
turned on~(\ref{WLBF}). From the worldsheet point of view, this means
that we start with a sphere and cut out $ g+1$ boundaries
representing the magnetised D-branes. Thus we are dealing with a
Riemann surface of genus zero, with $g+1$ borders and no crosscaps; in
the open string channel it corresponds to a $g$-loop diagram. 

We start from the result obtained in~\cite{Russo:2007tc} for D-branes
without multiple wrappings and where all cocycle phases were
ignored\footnote{We follow as much as possible the conventions of that
  paper.}. In the open string channel, one of the D-branes (whose
identification matrix is indicated with $R_0$) is singled out as the
external border of the diagram; thus it is natural to introduce the
monodromy matrices ${\cal S}_\mu \equiv R_{0}^{-1} R_{\mu}$, with
$\mu=1,\ldots,g$, whose eigenvalues are $\ex{\pm
  2\pi\ii\e_\mu^\alpha}$
~($\alpha=1,2,..,d$).
The only assumption we will make on the monodromy matrices
$\mathcal{S}_{\mu}$ is that they commute with each other, namely that
\begin{equation}\label{coms}
\left[ \mathcal{S}_{\mu},\mathcal{S}_{\nu} \right]=0~.
\end{equation}
Notice that this does not imply that the identification matrices $R_i$
with $i=0,\dots,g$ also commute with each other. Of course the
converse holds and~\eq{coms} is implied by the requirement that
$\left[R_i,R_j \right]=0$. By following the classification
of~\cite{Bianchi:2005yz} this more restrictive constraint is related
to configurations with parallel magnetic fluxes. However, while
Eq.~\eq{coms} is invariant under the T-Duality, the constraint among
the identification matrices $R$  is not, as they do not transform by
a similarity transformation (see Eq.~\eq{Fduality}).
 Thus we will consider the slightly more
general class of configurations satisfying~\eq{coms}. In this case, it
is convenient to perform a T-Duality and transform the zero-th D-Brane
into a purely Dirichlet D-Brane, i.e. with $R_0=-1$. As discussed in
the previous section, this can be done with a T-duality having $c^{-1}
d=F_0$.  With this choice we have $\mathcal{S}_{\mu}=-R_{\mu}$ and the
commutator above can be rewritten as
\begin{equation}
  \label{parallel}
  \left[ G^{-1} {\cal F}_{\mu},G^{-1} {\cal F}_{\nu} \right]=0~,
\end{equation}
with $\mu=1,\ldots,g$ only. This implies that there exists a complex
basis in which all the $G^{-1}\mathcal{F}_{\mu}$ are diagonal as in
Eq.~(\ref{F'}).
Moreover it implies: 
\begin{equation}
\label{comF}
  \left[ G^{-1}(F_{\hat\mu}-F_g),G^{-1}(F_{\hat\nu}-F_g)  \right]=0,~
~~~~~~\forall {\hat\mu},{\hat\nu}=1,\dots,g-1~;
\end{equation}
as a consequence, for a generic $G$ one can deduce that a fundamental
cell of the lattice torus exists where all the field differences 
$(F_{\hat\mu}-F_g)$ are simultaneously block diagonal.
 At any time, we can use again the T-duality rules
discussed in the previous section and go back to the original system
with all magnetized D-branes.

We start from Eq.~(3.29) of~\cite{Russo:2007tc} describing the
$g$-loop partition function in the open string channel
\begin{equation} \label{opchT} 
Z_g (F) \sim \left[ \prod_{\mu = 1}^{g} \sqrt{{\rm
 Det}\left(1-G^{-1} {\cal F}_\mu\right)}\right]
\int\left[d Z \right]_g \;\mathcal{A}^{(0)}\;
\prod_{\alpha=1}^d \left[ \ex{- \ii \pi \vec{\e^\alpha} \cdot \tau \cdot
\vec{\e}^\alpha} \; \frac{\det \tau }{\det
T_{\vec{\e}^\alpha} } \; {\cal R}_g \left(
  \vec{\e}^\alpha \cdot \tau \right) \right]\,,
\end{equation}
where $\left[d Z \right]_g$ is the untwisted ($F_i=0$) result and
$\vec{\e}^\alpha$ collects in a vector of length $g$ all the twists
${\e}^\alpha$; $\tau$ and $T_{\vec{\e}^\alpha}$ are the standard and the
twisted period matrix respectively.~~~
 $\mathcal{A}^{(0)}$ is the
classical contribution to the partition function calculated in
\cite{Russo:2007tc} in absence of the cocycle phases 
and setting all of the D-branes wrappings to one:
\begin{equation}\label{claw1} 
\mathcal{A}^{(0)} = 
\sum\, \Delta\; 
\exp\!\left\{\!\pi\ii \sum_{\hat\mu,\hat\nu=1}^{g-1} \a_0^{\hat\mu} G {\cal
    S}_{\hat\mu}^{-1/2} \bm{D}_{\hat{\mu}\hat{\nu}}{\cal S}_{\hat{\nu}}^{1/2}
  ~\a_0^{\hat\nu}\right\} ~,
\end{equation}
where $\bm{D}_{\hat{\mu}\hat{\nu}}$ is a space-time matrix determined
by the ${\cal S}_\mu$'s and the sum is over all the winding numbers
that satisfy the Kronecker's deltas representing the
identification~\eq{nFm} for each boundary state and the Kaluza-Klein
and winding conservations (recall that closed strings emitted by the
D-Brane with $R_0=-1$ are characterized by unconstrained Kaluza-Klein
momenta and no winding numbers):
\begin{equation}
  \label{Kdelta}
  \Delta = \left[
    \prod_{\mu=1}^g \delta\left(\hat{n}_{\mu}+
   F_{\mu} \hat{m}_{\mu}\right)\right]
  \delta\left(\sum_{i=0}^g \hat{n}_i\right) 
  \delta\left(\sum_{\mu=1}^g \hat{m}_{\mu}\right)~.
\end{equation}
All the $\e$-dependent ingredients, including the function ${\cal
  R}_g$ and the matrix $\bm{D}$, are defined in~\cite{Russo:2007tc}
and, of course, depend on the moduli of the Riemann surface. We do
not need the precise form of all these ingredients, but only some
properties that we will recall later in this section. Notice that the
classical contribution~\eq{claw1} is nontrivial only for $g \geq 2$:
for the annulus we have $\mathcal{A}^{(0)} = 1$ and Eq.~\eq{opchT}
reduces to the partition function in the uncompact
space~\cite{Bachas:1992bh}.

In this section we complete~\eq{claw1} to include also the effects of
the cocycle phases, multiple wrappings and open string moduli (Wilson
lines and/or D-brane positions). The classical contribution is the
only part of the partition function that is affected by this
generalization, as it is clear from the form of the interaction
vertex~(\ref{Nvertex}) and of the magnetized D-Branes boundary
states~(\ref{WLBF}). Basically we need to include in the sewing
procedure the cocycle factor in~(\ref{Nvertex}) and the
phases~
(\ref{WLBF}). Of course by
following this approach we are effectively working in the closed
string description. However the new contribution is independent of the
world-sheet moduli and thus can be directly included in Eq.~\eq{opchT}
which is written in the open string channel. It is then clear that,
in the expression for $\mathcal{A}$, we will have the same exponential
of Eq.~\eq{claw1} multiplied by some additional factors related to
cocycles. So let us consider these new contributions: by using~\eq{Nvertex},
~\eq{nFm} and~\eq{WLBF}, one can see that all the phases from the cocycles and
the Wilson lines yield
\begin{eqnarray}\label{coco}
&& \mathrm{exp} \left\{ 2\pi \ii \left[ \sum_{\mu=1}^{g}
    C_{\mu} \hat{m}_{\mu} +Y_0\hat{n}_0 \right] \right\} 
  \times   \exp  \left\{ \pi \ii \sum_{\mu=1}^{g}
    \sum_{M<N}\hat{m}_{\mu}^M (F_{\mu})_{MN}\, \hat{m}_{\mu}^N \right\}\\ 
 & \times & \nonumber \exp\left\{ \frac{\pi \ii}{2} \sum_{\nu>\mu=1}^{g}
    \left(\hat{m}_\mu F_\mu \hat{m}_\nu + \hat{m}_\mu
    F_\nu \hat{m}_\nu \right) 
 \right\} ~.
\end{eqnarray}
where $Y_0$ encodes the position of the zero-th D-Brane which is
point-like along the torus directions. By using~\eq{Kdelta}, we can
eliminate $\hat m_g$ from these sums.  Then it is easy to see that we
can rewrite the dependence on the open string moduli as follows:
\begin{equation}\label{wld}
  \exp\left\{2\pi\ii\sum_{\hat{\mu}=1}^{g-1} \left[ C_{\hat{\mu}}- C_{g}+
Y_0(F_{\hat{\mu}}-F_g) \right]
      \hat{m}_{\hat{\mu}}\right\} \equiv
    \ex{2\pi\ii \sum\limits_{\hat{\mu}=1}^{g-1} {}^t\!\rho_{\hat{\mu}}
      \hat{m}_{\hat{\mu}}}\,.
\end{equation}
The second exponential in~\eq{coco} comes from the boundary states;
using the conservation of the winding numbers, the exponent can be
 rewritten as follows:
\begin{equation}
  \sum_{\mu=1}^g\sum_{M<N} \hat{m}_{\mu} ^M (F_{\mu})_{MN} 
  \hat{m}_{\mu}^N =  \sum_{\hat{\mu}=1}^{g-1}\sum_{M<N} 
  \hat{m}_{\hat{\mu}}^M (F_{\hat{\mu}})_{MN} \hat{m}_{\hat{\mu}}^N
  - \sum_{\hat{\mu},\hat{\nu}=1}^{g-1} \sum_{M<N}
  \hat{m}_{\hat{\mu}}^M (F_g)_{MN} \hat{m}_{\hat{\nu}}^N~.
\end{equation}
 By combining this contribution with the 
last exponent in Eq.~\eq{coco} and using~\eq{Kdelta} we get
\begin{equation}\label{e1final}
  \exp\left[-\pi\ii \sum_{\hat{\mu},\hat{\nu}=1}^{g-1} 
  \sum_{M<N} \hat m_{\hat{\mu}}^M
    (F_{\hat{\mu}\hat{\nu}})_{MN} \hat m_\nu^N \right] ~,
\end{equation} 
 where 
\begin{equation}\label{Fmunu}
F_{\hat{\mu}\hat{\nu}}= F_{\hat{\nu}\hat{\mu}}=F_{\hat{\nu}}-F_g 
~~,~~~\mbox{for}~\hat\nu\geqslant\hat\mu.
\end{equation} 
Observe that in order to obtain this final form we have changed the sign of the combination $\sum_{M<N}\hat{m}^M_{\mu}(F_{\mu})_{MN}\hat{m}^N_{\mu}$ as each term of the sum is integer (see the Appendix).

Thus the total contribution from the
various phase factors is just the product of~\eq{wld}
and~\eq{e1final}. This expression has no dependence on the metric of
the torus and, in particular, Eq.~\eq{e1final} provides just some
relative signs between contributions related to different values of
$\hat{m}$.
Moreover it depends only on the differences $(F_{\hat{\nu}}-F_g)$;
therefore, thanks to Eq.~\eq{comF}, we can always consider 
$F_{\hat{\mu}\hat{\nu}}$ block diagonal.

Let us now reconsider the exponential in~\eq{claw1}. By
using~\eq{zeromodes}, \eq{nFm}, \eq{parallel} and the various
conservation laws we can rewrite it as follows
\begin{equation}
  \label{erew}
  \exp \left\{ \frac{\pi \ii}{2}
    \sum_{\hat{\mu},\hat{\nu}=1}^{g-1}{}^t\hat{m}_{\hat{\mu}}
    G\left[ 1- \left( G^{-1}\mathcal{F}_{\hat{\mu}}\right)^2
    \right]^{\frac{1}{2}} 
    \bm{D}_{\hat{\mu}\hat{\nu}}(\mathcal{S}) 
   \left[ 1- \left(
      G^{-1}\mathcal{F}_{\hat{\nu}} \right)^2
  \right]^{\frac{1}{2}} \hat{m}_{\hat{\nu}} \right\}~,
\end{equation}
where we explicitly remind that $\bm{D}$ is a function of the
space-time matrices ${\cal S}$. It is convenient to rewrite~\eq{erew}
in the complex basis defined by the vielbein $\mathcal{E}$, where the
$G^{-1}\mathcal{F}_i$'s are diagonal and denoted by
$\mathcal{F}^{(d)}_i$ as in Eq.~(\ref{F'}):
\begin{equation}\label{erew2}
\mathrm{exp}\left\{ \frac{\pi\ii}{2} \sum_{\hat{\mu},\hat{\nu}=1}^{g-1}
  {}^t\hat{m}_{\hat{\mu}} \left[{}^t\!\mathcal{E}
  \mathcal{G}\sqrt{\left( 1-(\mathcal{F}_{\hat{\mu}}^{(d)})^{2} \right)
   \left( 1-(\mathcal{F}^{(d)}_{\hat{\nu}})^2 \right)} \left( 
 \begin{array}{cc}
\mathcal{D}_{\hat{\mu}\hat{\nu}}(\epsilon) & 0\\
0 & \mathcal{D}_{\hat{\mu}\hat{\nu}}(-\epsilon)
\end{array} \right)\mathcal{E}\right]
\hat{m}_{\hat{\nu}}\right\}~,
\end{equation}
where now each $\mathcal{D}_{\hat{\mu}\hat{\nu}}$ is $d\times d$
diagonal matrix that depends on the eigenvalues of the ${\cal
  S}_\mu$'s. 

The square
parenthesis in~\eq{erew2} is contracted with a symmetric combination
of $\hat{m}$, so we can symmetrize it. Then, by using~\eq{offmetric}, one
can easily check that~\eq{erew2} is equal to
\begin{equation}\label{erew3}
\mathrm{exp}\left\{ \frac{\pi\ii}{2} \sum_{\hat{\mu},\hat{\nu}=1}^{g-1}
  {}^t\hat{m}_{\hat{\mu}} \left[{}^t\!\mathcal{E}
  \mathcal{G}\sqrt{\left( 1-(\mathcal{F}^{(d)}_{\hat{\mu}})^2 \right)
   \left( 1-(\mathcal{F}^{(d)}_{\hat{\nu}})^2 \right)} \left( 
 \begin{array}{cc}
\hat{\tau}_{\hat{\mu}\hat{\nu}} & 0\\
0 & \hat{\tau}_{\hat{\nu}\hat{\mu}}
\end{array} \right)\mathcal{E}\right]
\hat{m}_{\hat{\nu}}\right\}~,
\end{equation}
where the $d\times d$ diagonal matrix\footnote{Our
  $\hat{\tau}$ is equal to the $\tau$ of~\cite{Antoniadis:2005sd}; we
  indicate it with a different symbol in order to avoid confusion with
  the standard period matrix.} 
 $\hat\tau_{\hat{\mu}\hat{\nu}}$
is given by:
\begin{equation}
  \label{hattau}
   \hat\tau_{\hat{\mu}\hat{\nu}} \equiv \frac 12 \left[
\mathcal{D}_{\hat{\mu}\hat{\nu}}(\e) +
\mathcal{D}_{\hat{\nu}\hat{\mu}}(-\e)\right]~.
\end{equation}

Expressing $\mathcal{E}$ in terms of the real vielbein $E$ 
($\mathcal{E}= S E$, with $S$ given in Eq.~\eq{Eps}) we can go to the
real basis,
where Eq.~\eq{erew3} reads:
\begin{equation}\label{erew3'}
\mathrm{exp}\left\{ \frac{\pi\ii}{2} \sum_{\hat{\mu},\hat{\nu}=1}^{g-1}
  {}^t\hat{m}_{\hat{\mu}} \left[{}^t\!E
  \sqrt{\left( 1-(\mathcal{F}^{(d)}_{\hat{\mu}})^2 \right)
   \left( 1-(\mathcal{F}^{(d)}_{\hat{\nu}})^2 \right)} \left( 
 \begin{array}{cc}
\hat{\tau}^S & \ii\hat{\tau}^A\\
 -\ii\hat{\tau}^A& \hat{\tau}^S
\end{array} \right)_{\hat{\mu}\hat{\nu}} E\right]
\hat{m}_{\hat{\nu}}\right\}~,
\end{equation}
where  $\hat\tau^S$ and $\hat{\tau}^A$ are the symmetric
 and the antisymmetric  part of  $\hat{\tau}$, in the
exchange of $\hat{\mu},\;\hat{\nu}$.
As  $\hat{\tau}$ is purely imaginary and  $Im~ \hat{\tau}^S$ is
positive definite because of the
Riemann bilinear identities~\cite{Antoniadis:2005sd} and moreover
$\left( \sqrt{ 1-(\mathcal{F}^{(d)}_{\hat{\mu}})^2} \right)_{ab}= 
\delta_{ab}~|\left(1-\mathcal{F}^{(d)}_{\hat{\mu}}\right)_{aa}|$, 
the convergence of the series in Eq.~\eq{claw1} is assured.

Following \cite{Bianchi:2005sa} we can rewrite the Born-Infeld square
roots above in yet another way by 
using another important consequence of the
Riemann bilinear identities~\cite{Antoniadis:2005sd}:
\begin{equation}\label{C}
C_{\hat{\mu}\hat{\nu}} =C_{\hat{\nu}\hat{\mu}}  \equiv 
\frac{1}{2}\left[ \mathcal{D}_{\hat{\mu}\hat{\nu}}(\epsilon)-
 \mathcal{D}_{\hat{\nu}\hat{\mu}}(-\epsilon)\right] =
  \ii \frac{\sin(\pi \epsilon_{\hat{\mu}})
  \sin(\pi \epsilon_{\hat{\nu}}-\pi\epsilon_{g})}
 {\sin(\pi \epsilon_g)},\,\,\,\,\,\, 
 \hat{\nu}\geqslant \hat{\mu}~,
\end{equation}
where also $C_{\hat{\mu}\hat{\nu}}$ and the sines are $d\times d$
diagonal matrices whose entries depend on the different
values of $\e$.  This will be useful for the analysis of the
degeneration limit in the next section. The $2d\times 2d$ matrix~
$\mathrm{diag}\left\{ \sin(\pi \epsilon_{\mu}),\sin(\pi\epsilon_{\mu})
\right\}$~ can be written as\footnote{We will not keep track of the
  sign choices for the square roots: they clearly depend on whether
  each $\e_\mu$ is negative or positive.}:
\begin{equation} \label{tru}
\left(
\begin{array}{cc}
\sin(\pi \epsilon_{\mu}) & 0\\
0 & \sin(\pi \epsilon_{\mu})
\end{array} \right) 
=\sqrt{\frac{1}{1-(\mathcal{F}^{(d)}_{\mu})^2}}~.
\end{equation}
This can be checked by rewriting the sine in terms of exponentials
which are directly related to the components of ${\cal S}$ in the
complex basis: $\sin(\pi \epsilon^\a_{\mu})=[\sqrt{2-{\cal
    S}_\mu^{-1}-{\cal S}_\mu}\,]_{\a\a}/2,~~\a=1,2,..,d $. Also, using the same
procedure, we have
\begin{equation}\label{tru1}
\left(
\begin{array}{cc}
\sin( \pi \epsilon_{\hat{\nu}}-\pi \epsilon_g) & 0\\
0 & \sin( \pi \epsilon_{\hat{\nu}}-\pi \epsilon_g)
\end{array} \right)=\frac{ \left(
\begin{array}{cc}
\ii & 0\\
0 & -\ii
\end{array} \right)
(\mathcal{F}^{(d)}_g-\mathcal{F}^{(d)}_{\hat{\nu}})}
{\sqrt{\left( 1-(\mathcal{F}^{(d)}_g)^2\right)
\left( 1-(\mathcal{F}^{(d)}_{\hat{\nu}})^2 \right)}}~,
\end{equation} 
with $\mathcal{F}^{(d)}_{\mu}$ defined as in Eq.~(\ref{F'}). From~\eq{tru}
and~\eq{tru1} one can see that
\begin{equation}\label{Cfinal}
C_{\hat{\mu}\hat{\nu}}=C_{\hat{\nu}\hat{\mu}}= \frac{ \left(
\begin{array}{cc}
1 & 0\\
0 & -1
\end{array} \right)
\left( \mathcal{F}^{(d)}_{\hat{\nu}}-\mathcal{F}^{(d)}_g\right)}
{\sqrt{\left( 1-(\mathcal{F}^{(d)}_{\hat{\mu}})^2 \right)
\left( 1-(\mathcal{F}^{(d)}_{\hat{\nu}})^2\right)}}~~,
\,\,\,\,\,\,\mbox{for}~~~
\hat{\nu}\geqslant \hat{\mu}~.
\end{equation}
Thus we can eliminate the square roots in~\eq{erew3} in favor of $C$;
then it is convenient to decompose $\hat\tau$ into its symmetric
($\hat{\tau}^S$) and the antisymmetric ($\hat{\tau}^A$) parts.
Hence we can use Eq.~(\ref{F'}) to rewrite the 
diagonal fields $ \mathcal{F}^{(d)}$ 
in terms of the $\mathcal{F}$'s, take advantage of the identity 
$\mathcal{F}_\mu-\mathcal{F}_\nu = F_\mu-F_\nu $, insert the
contribution of the cocycles given by Eq.~\eq{e1final}  and finally get the
classical contribution to the twisted partition function
describing wrapped D-branes on an generic $T^{2d}$ :
\begin{eqnarray}
  \mathcal{A} & =& \sum \Delta \;
\mathrm{exp}\left\{{\pi\ii}
  \sum_{\hat{\mu},\hat{\nu}=1}^{g-1} \sum_{M<N}
  {}^t\hat{m}_{\hat{\mu}}^M  (F_{\hat{\mu}\hat{\nu}})_{MN}
   \hat{m}_{\hat{\nu}}^N\right\}\;
 \ex{2\pi \ii \sum\limits_{\hat{\mu}=1}^{g-1} {}^t\!\rho_{\hat\mu} \hat
   m_{\hat\mu}}
\label{awv1} \\ \nonumber &\times&
\mathrm{exp}
\left\{\frac{\pi\ii}{2} \sum_{\hat{\mu},\hat{\nu}=1}^{g-1}
{}^t\hat{m}_{\hat{\mu}} 
{}^t\mathcal{E} \left[ \left(
\begin{array}{cc}
\frac{\hat\tau^S}{C} & 0\\
0 & -\frac{\hat\tau^S}{C}
\end{array} \right)_{\hat{\mu}\hat{\nu}} \!\!\! +
\left(
\begin{array}{cc}
\frac{\hat\tau^A}{C} & 0\\
0 & \frac{\hat\tau^A}{C}
\end{array} \right)_{\hat{\mu}\hat{\nu}} 
\right] {}^t\mathcal{E}^{-1} F_{\hat{\mu}\hat{\nu}}
\hat{m}_{\hat{\nu}} \right\}~.
\end{eqnarray}
where $ F_{\hat{\mu}\hat{\nu}}$ is the matrix defined in
Eq.~\eq{Fmunu}\footnote{The inverse of $C$ must be meant only with
  respect to the Lorentz indeces, at fixed $\hat{\mu}$ and
  $\hat{\nu}$.}.  Then the full partition function is simply given
by~\eq{opchT}, where $\mathcal{A}^{(0)}$ is substituted by the
complete expression above.

If we restrict ourselves to the case of a factorizable torus
$T^{2d}=(T^2)^d$, then~\eq{awv1} agrees\footnote{Apart from some
  factors of two.} with the results of~\cite{Antoniadis:2005sd}. In
order to make contact with their setup it is first useful to perform a
T-Duality in such a way that the singled-out boundary with
identifications encoded in the $R_0$ matrix is transformed back to a
magnetized D-Brane. From~\eq{Fduality} we can compute the
transformation of the difference between two gauge field strengths
\begin{equation}
  \label{fdiff}
  F'_i-F'_j = (F_j\,{}^t c + \,{}^t d)^{-1} (F_j-F_i) (cF_i-d)^{-1}~.
\end{equation}
In particular the duality we performed to put $R_0=-1$ had $d=cF_0$.
Then by combining Eq.~(\ref{fdiff}) with the transformation of the
winding numbers (\ref{n'm'}), the amplitude~\eq{awv1} reduces to the
following product of terms, each one related to a single $T^2$:
\begin{eqnarray}
  \mathcal{A}^w_{(T^2)^d} \!\!\!& =& \! \sum \Delta \;
\mathrm{exp}\left\{{\pi\ii}
  \sum_{\hat{\mu},\hat{\nu}=1}^{g-1} \sum_{M<N}
  {}^t\hat{m}_{\hat{\mu}}^M (F_{\hat\mu\hat\nu})_{MN}
   \hat{m}_{\hat{\nu}}^N\right\}\;
 \ex{2\pi \ii \sum\limits_{\hat{\mu}=1}^{g-1} {}^t\!\rho_{\hat\mu} \hat
   m_{\hat\mu}}
\nonumber  \\ &\times& \label{t2fact}
\prod_{\a=1}^d\mathrm{exp}\left\{\frac{\pi}{2}
\sum_{\hat{\mu},\hat{\nu}=1}^{g-1} {}^t \hat m^\a_{\hat{\mu}}\left[
  \left(\frac{\tau_S(\epsilon^\a)}{C(\epsilon^\a)}\right)_{\hat{\mu}\hat\nu}
  \!\! \mathcal{I}^{(\a)} +\ii \left(
  \frac{\tau_A(\epsilon^\a)}{C(\epsilon^\a)}\right)_{\hat{\mu}\hat\nu}\right]
  F^{(\a)}_{\hat{\mu}\hat\nu} \hat{m}^\a_{\hat\nu} \right\} ~,
\end{eqnarray}    
where now
\begin{equation}
F_{\hat{\mu}\hat{\nu}}=F_{\hat{\nu}\hat{\mu}}=(F_0-F_{\hat{\mu}})(F_0-F_g)^{-1}
(F_{\hat{\nu}}-F_g)
~~,~~~\mbox{for}~\hat\nu\geqslant\hat\mu,
\end{equation} 
and $\mathcal{I}^{(\a)}$ is the complex structure of each
of the $T^2$'s defined as Eq.~(\ref{RealComplexStructure}). This can
be compared with Eq.~(A.28) of~\cite{Antoniadis:2005sd}. The norm of
the vector $v_i$ appearing there is related to the Born-Infeld square
roots: ${|v_i U|}/{\sqrt{U_2 T_2}} = w_{\hat{\mu}} \sqrt{\left(
    1-(\mathcal{F}^{(d)}_{\hat{\mu}})^2 \right)}$. One can use this
in~\eq{Cfinal} and~\eq{t2f} together with the explicit expression for
the $T^2$ complex structure~\eq{cst2} to check that the two results
are related by a further T-duality that exchanges $T\leftrightarrow
-1/U$.

\sect{Twist fields couplings on $T^{2d}$.} \label{tft2d}

In this section we will focus on the vacuum diagram with three
boundaries (i.e. $g=2$) and study the degeneration limit where all
three open string propagators in Fig.~\ref{2l}b become long and thin.
In this situation the partition function factorizes in two tree-level
3-point correlators between twist fields. This result provides the
main contribution for the computation of the Yukawa couplings in
string phenomenological models and of other terms in the effective
action generated by stringy
instantons~\cite{Blumenhagen:2006xt,Ibanez:2006da}. For $g=2$ the only
non-vanishing entry for $\hat\tau$ is clearly $\hat\tau^S_{11}$, which
has been computed, at leading order in this degeneration limit,
in~\cite{Russo:2007tc}. Again the novelty of the present computation
resides in the analysis of the classical part~\eq{awv1}, while all
other ingredients of the partition function~\eq{opchT} factorize
exactly as before, since they do not depend on the wrapping numbers or
the Wilson lines. In the degeneration limit under study we
have~\cite{Russo:2007tc} that $\mathcal{D}_{11}(\epsilon) \rightarrow
0$ from which
\begin{equation}
\left( \frac{\hat{\tau}^s(\epsilon)}{C(\epsilon)} \right)_{11} 
\rightarrow -1_{d\times d}
\end{equation}
thus the exponential in the second line of Eq.~(\ref{awv1}) becomes
\begin{equation}\label{e2finalg2}
\mathrm{exp}\left\{\frac{\pi}{2}{}^t \hat{m}_1 \mathcal{I}(F_2-F_1)
\hat{m}_1 \right\}~,
\end{equation}
where we have introduced the complex structure of the torus in the
integral basis as in Eq.~(\ref{RealComplexStructure}). In order to
write the final form of the amplitude as a sum over unconstrained
integers, it is necessary to solve the conservations (\ref{Kdelta}),
which, in the case under study, become
\begin{equation}
\hat{n}_1=-F_1\hat{m}_1\,\,\,,\,\,\,\hat{n}_2=-F_2\hat{m}_2\,\,\,,\,\,\,
\hat{m}_1+\hat{m}_2=0~.
\end{equation}
Of course the solutions must have integer Kaluza-Klein and windings
numbers, so there must exist a minimal\footnote{By minimal we mean
  that any other matrix with the same property is an integer
  multiple of $H$.} integer invertible matrix $H$ such that $F_1 H$
and $F_2 H$ are integer matrices and the solution can be written as
$\hat{m}_1=Hh$, with $h \in \mathbb{Z}^{2d}$.  Then we define
$\mathcal{I}'={}^t\!H\mathcal{I}\,{}^t\!H^{-1}$, which is still a
complex structure, and $F\equiv {}^t\!H(F_2-F_1)H$; so the
degeneration limit of the amplitude (\ref{awv1}) in our $g=2$ case is
\begin{equation}\label{Azm2}
\mathcal{A}= \sum_{h \in \mathbb{Z}^{2d}} \mathrm{exp}  
\left\{ \frac{\pi}{2}\,\left[{}^t\!h \mathcal{I}'F h+
\ii \sum_{M<N} h^M F_{MN} h^N \right] \right\}\times  \mathrm{exp}  
\left\{ 2\pi \ii \,{}^t\!\rho_1 Hh \right\} ~,
\end{equation}
with a possible half-integer shift of $\rho_1$ (see the Appendix for
further details). 

By unitarity it must be possible to rewrite the result in~(\ref{Azm2})
as a sum where each term is the product of two functions that are one
the complex conjugate of the other. Each function represents the
classical contribution to the coupling among three twist fields, while
the quantum contribution follows from the factorization of the other
terms in the partition functions~\eq{opchT}, see~\cite{Russo:2007tc}.
The presence of the sum is due to the fact that the vacuum describing an
open string stretched between two magnetized D-branes has a finite
degeneracy~\cite{Abouelsaood:1986gd} in compact spaces. So each string
state has a number of replica and the various terms in the sum
describe the couplings between these different copies of the twist
fields (of course this is exactly what we need, in phenomenological
models, to describe the Yukawa couplings for different families).  
Let us see how this works in the simple case of the 2-torus.

\subsection{The two-torus example}

For a generic tilted torus the metric can be written as a function of
the K\"ahler and complex structure moduli
\begin{equation}\label{G}
G=\frac{T_2}{U_2} \left(
\begin{array}{cc}
1 & U_1\\
U_1 & |U|^2
\end{array} \right)
={}^t \mathcal{E} \left(
\begin{array}{cc}
0 & 1\\
1 & 0 
\end{array} \right)
\mathcal{E},
\end{equation}  
having defined the complex structure as $U=U_1+\ii U_2$ and the
K\"ahler form as $T=T_1+\ii T_2$. Thus the vielbein (\ref{offmetric})
reads
\begin{equation}\label{t2v}
\mathcal{E}=\sqrt{\frac{T_2}{2U_2}}\left(
\begin{array}{cc}
1 & U\\
1 & \bar{U}
\end{array} \right)
\end{equation}
from which one can write the explicit form of the $T^2$ complex
structure in the integral basis
\begin{equation}\label{cst2}
\mathcal{I}={}^t \mathcal{E} \left(
\begin{array}{cc}
\ii & 0\\
0 & -\ii
\end{array} \right)
{}^t \mathcal{E}^{-1}=-\frac{1}{U_2}\left(
\begin{array}{cc}
U_1 & -1\\
|U|^2 & -U_1
\end{array} \right),
\end{equation}
The magnetic fields on the two magnetized D-Branes are identified by the 
Chern numbers $p_i$ and the products of the wrappings along the
two cycles of the torus $W_i$: 
\begin{equation}\label{t2f}
F_i= \left(
\begin{array}{cc}
0 & \frac{p_i}{W_i}\\
-\frac{p_i}{W_i} & 0
\end{array} \right),~~~~i=1,2.
\end{equation}
One can easily check that the matrix $H$ is simply proportional to the
$2 \times 2$ identity, namely $H=W_1W_2/\delta \otimes 1_{2 \times
  2}$, where $\delta=\mathrm{G.C.D}\left\{ W_1,W_2\right\}$. This
implies that Eq.~(\ref{Azm2}) can be put in the following form
\footnote{In the configuration of \cite{Russo:2007tc} one gets $I>0$,
  otherwise one should write $|I|$, because of the note before
  Eq.~\eq{tru}.}
\begin{equation}\label{n422}
\sum_{h_1,h_2 \in \mathbb{Z}}\exp \left\{ -\frac{\pi}{2}\frac{I}{U_2}
\left[ h_1^2+|U|^2h_2^2+2Uh_1h_2 \right] + 
2\pi \ii \frac{1}{\delta}\mathcal{C}_M h^M \right\},
\end{equation}
with
\begin{equation}\label{I}
I=\frac{W_1^2W_2^2}{\delta^2}
\left(\frac{p_2}{W_2}-\frac{p_1}{W_1}\right)=
I_{21}\frac{W_1W_2}{\delta^2} ~,
\end{equation}
where we introduced the intersection numbers
\begin{equation}\label{Iij}
I_{ij}=p_iW_j-p_jW_i
\end{equation}
and 
\begin{equation}
\mathcal{C}_M=W_1W_2\left[(F_1-F_2)Y_0+(C_1-C_2)\right]_M~.
\end{equation}
We want to perform a T-Duality in such a way that also the zero-th D-Brane
is  magnetized. The intersection numbers are invariant under this
operation, while the form of the matrix $I$ is modified due to the
transformation of the wrapping numbers. Recalling that the T-Duality
relating a Dirichlet to a magnetized brane is encoded in an
$O(2,2,\mathbb{Z})$ matrix of the type in Eq.~(\ref{Tduality})
with\footnote{We indicate with a tilde the quantities in the picture
  with a magnetized zero-th D-Brane.}
\begin{equation}\label{zeroDuality}
d=\left(
\begin{array}{cc}
0 & -p_0\\
p_0 & 0
\end{array} \right)\,\,\,\,\mathrm{and}\,\,\,\,c=\tilde{w}_0 \equiv
\tilde{W}_0 \otimes 1_{2\times 2}~, 
\end{equation}
it is possible to show, combining the invariance of $I_{21}$ with
Eq.~(\ref{fdiff}), that $W_1 \rightarrow I_{01}$ and $W_2 \rightarrow
I_{20}$. Thus
\begin{equation}\label{I'} 
I 
=\frac{I_{01}I_{21}I_{20}}{\delta^2}
\end{equation}
with $\delta=\mathrm{G.C.D.}\left\{ I_{20},I_{10}\right\}
=\mathrm{G.C.D.}\left\{ I_{20},I_{10},I_{21}\right\}$, since we can make
use of the property
$I_{21}\tilde{W}_0+I_{20}\tilde{W}_1+I_{01}\tilde{W}_2=0$. Under the
same duality the open string moduli transform as follows
\begin{equation}
  C_{\mu}=\frac{\tilde{C}_{\mu}}{\tilde{w}_0(F_{0}-F_{\mu})}
\,\,\,\,\,\,\mathrm{and}  \,\,\,\,\,\,Y_0=\tilde{w}_0\tilde{C}_0 
\end{equation}
hence
\begin{equation}
{\mathcal{C}}_M=
\tilde{W}_0I_{12}\tilde{C}^{(0)}_M+\tilde{W}_1I_{20}\tilde{C}^{(1)}_M+ 
\tilde{W}_2I_{01}\tilde{C}^{(2)}_M.
\end{equation}
where the superscript $(i)$ indicates the three boundaries and the
subscript $M$ is the Lorentz index. 

The configurations studied~\cite{Russo:2007tc} had $\delta$ and all
$\tilde{W}_i$ equal to $1$. In this case $I$ is always an even number,
since it is the product of three integers that sum to zero. Thus the
contribution of the cocycles in~\eq{Azm2} is irrelevant and~\eq{n422}
can be rewritten as it was done in~\cite{Russo:2007tc}. We choose not
to do that here, because it is easier to deal always with~\eq{n422}
without treating the case $\tilde{W}_i=1$ separately.  In order to
factorize the amplitude above
and find the Yukawa couplings corresponding to the states of the open
strings stretched between pairs of D-Branes with different magnetic
fields on their world volume, it is necessary to first perform a
Poisson resummation on the integer $h_1$
\begin{equation}
  \sum_{h_1=-\infty}^{+\infty}\ex{-\pi Ah_1^2+2\pi
    h^1As}=\frac{1}{\sqrt{A}}
  \ex{\pi As^2}\sum_{h_1=-\infty}^{+\infty}
  \ex{-\pi \frac{h_1^2}{A}-2\pi \ii h_1s}~,~~~~~A>0.
\end{equation}
This yields a new form of the same amplitude which is easy to
factorize once we introduce a new pair of integers, $r$ and $k$,
through the relations
\begin{equation}
h_1= rI+l =I \left( r+\frac{l}{I} \right) 
\,\,\,\,\,\,\mathrm{and}\,\,\,\,\,\, h_2=k-r~,
\end{equation}
where $l=1,\dots,I$. It is manifest that in this way both the former
and the latter pair of integers range in the whole $\mathbb{Z}$.
Notice that this is ensured by summing over the additional integer $l$
as well. Simple algebraic manipulations then lead to the product of
two Jacobi Theta-functions as follows
\begin{equation}
\mathcal{A}=\sqrt{\frac{2U_2}{I}}\sum_{l=1}^{I} \vartheta \left[
\begin{array}{c}
\frac{l}{I}-\frac{1}{I}\frac{\mathcal{C}_1}{\delta}\\
\frac{\mathcal{C}_2}{\delta}
\end{array} \right] \left( 0 \big| IU \right) \times \vartheta \left[
\begin{array}{c}
\frac{l}{I}-\frac{1}{I}\frac{\mathcal{C}_1}{\delta}\\
-\frac{\mathcal{C}_2}{\delta}
\end{array} \right] \left( 0 \big| -I\bar{U} \right)
\end{equation}
This result generalizes the one of ref.\cite{Russo:2007tc} and is in 
agreement\footnote{The apparent   mismatch related to
  the presence of the wrapping numbers of the D-Branes in the Wilson
  lines dependence of the couplings is resolved by checking that
  $W_iC^{(i)}_1\,,\,W_iC^{(i)}_2 \in [0,1]$ as well as the parameters
  $\epsilon_i$ and $\theta_i$ defined in \cite{Cremades:2003qj}.} with
the Section 3.1.3 of \cite{Cremades:2003qj}, as we find that, if the
intersection numbers $I_{ij}$ are not coprime, one has
$I_{20}I_{01}I_{21}/\delta^2$ non vanishing Yukawa couplings, labeled
by the integer $l$. Indeed, in the dual picture involving intersecting
D-Branes, every open string living in the intersection between two
fixed D-Branes, say for instance $i$ and $j$, will only couple to
$|I_{jk}I_{ik}|/\delta^2$ strings from the intersections between the
D-Brane $k$ and the D-Branes $i$ and $j$ \cite{Cremades:2003qj}.

\subsection{Twist fields couplings on a generic $T^{2d}$ compactification}

In order to factorize the classical contribution to the partition
function~(\ref{Azm2}) in the most general case analyzed in Section
\ref{SectionAmplitude} and read the corresponding twist field
couplings, we need to use the properties of the complex structure. Let
us first decompose ${\cal I}'$ in terms of its $d \times d$ blocks
\begin{equation}
\mathcal{I}'=\left(
\begin{array}{cc}
A & B\\
C & D
\end{array} \right)~.
\end{equation}
Hence it is simple to check that ${\cal I'}^2=-1$ yields
\begin{equation}
  \label{ABCD}
AB=-BD~~,~~~ \mbox{and}~~ A^2=-(1+BC)~.
\end{equation}
Then we choose a basis for the torus lattice where the combination
${}^t\!H(F_2-F_1)H$ takes the following form
\begin{equation}\label{Fblock1} 
F=\left(
\begin{array}{cc}
0 & \hat{F}\\
-\hat{F} & 0
\end{array} \right).
\end{equation}
This can be done by putting the matrix $F$ in the form of
Eq.~(\ref{Fblock}) (thanks again to the result of the Appendix) and by
a suitable relabeling of the rows and the columns. Notice that this
relabeling does not affect the form of the amplitude to be factorized.
For sake of brevity, we also introduce the $2d$-components vector
$\beta=\,{}^tH\rho_1$. Then the general amplitude to be factorized has
the following form\footnote{In the following $i$ indicates a $d\times
  d$ imaginary matrix: $i\equiv \ii\, 1_{d\times d}$.}
\begin{equation}\label{Azm3}
\sum_{h_i \in \mathbb{Z}^d} \exp \left\{ \frac{\pi}{2} 
\left[ \left( {}^t\!h_1\, {}^t\!h_2 \right) \left(
\begin{array}{cc}
-B\hat{F} & (i +A)\hat{F}\\
(i-D)\hat{F} & C\hat{F}
\end{array} \right) \left(
\begin{array}{c}
h_1\\
h_2
\end{array} \right) \right]+2\pi \ii \,
{}^t\!\beta_1h_1+2\pi \ii \,{}^t\!\beta_2 h_2 \right\}~.
\end{equation}
As next step we need to  perform a Poisson resummation 
\begin{equation}\label{Poisson}
\sum_{h_1 \in \mathbb{Z}^d}\ex{-\pi \,{}^t\!h_1\mathcal{B}h_1+
2\pi\,{}^t\!h_1\mathcal{B}s}=\frac{1}{\sqrt{\mathrm{Det}\mathcal{B}}}
\ex{\pi \,{}^t\!s\mathcal{B}s}\sum_{h_1 \in \mathbb{Z}^d}
\ex{-\pi \,{}^t\!h_1\mathcal{B}^{-1}h_1-2\pi \ii \,{}^t\!h_1s}
\end{equation}
on the first $d$ components of $h$ which we indicate with $h_1$. So in
our case we have
\begin{equation}
\mathcal{B}=\frac{1}{2} \hat{F}\,{}^t\!B \,\,\,\,\,\, 
\mathrm{and} \,\,\,\,\,\, 
s=\frac{1}{2}\mathcal{B}^{-1}(A+i)\hat{F}h_2+
\ii \mathcal{B}^{-1}\beta_1~.
\end{equation}  
The first exponential in the r.h.s. of the Poisson resummation formula
(\ref{Poisson}) yields a quadratic term in the vector $h_2$, which can
be combined with a similar contribution present in the initial
expression~(\ref{Azm3})
\begin{eqnarray}
\nonumber && \frac{\pi}{4} \,{}^t\!h_2 \hat{F} (i+\,{}^t\!A)\, 
{}^t\!\mathcal{B}^{-1}(i+A)\hat{F}h_2+
\frac{\pi}{2}\,{}^t\!h_2 C\hat{F} h_2=\\
&& =  \frac{\pi}{2}\,{}^t\!h_2 
\left[ \hat{F}(i+\,{}^t\!A)\hat{F}^{-1}B^{-1}(i+A)\hat{F}+
C\hat{F} \right]h_2~.
\end{eqnarray}    
Recalling that $\mathcal{I}'\hat{F}$ is a symmetric matrix, it is not difficult
to see that $A\hat{F}=-\hat{F}\,{}^t\!D$. Some algebraic
manipulations involving these identities simplify the previous
expression into
\begin{equation}
\ii \pi \,{}^t\!h_2 B^{-1}(i+A)\hat{F}h_2.
\end{equation}
Hence the Poisson resummation performed on Eq.~(\ref{Azm3}) gives
\begin{eqnarray}
\nonumber \frac{1}{\sqrt{\mathrm{Det}(2B\hat{F})}} &&
\sum_{h_i \in \mathbb{Z}^d} 
\exp \left\{ \ii \pi \left[ \,{}^t\!h_2B^{-1}(i+A)
\hat{F}h_2+2\ii \,{}^t\!\beta_1\,{}^t\!B^{-1}
\hat{F}^{-1}\beta_1+ \right. \right. \\
\nonumber && +\,{}^t\!h_2\hat{F}(i+\,{}^t\!A)\,{}^t\!
\hat{F}^{-1}B^{-1}\beta_1+\,{}^t\!\beta_1
\hat{F}^{-1}B^{-1}(i+A)\hat{F}h_2\\
\nonumber && +2\,{}^t\!\beta_2h_2+2\ii \,{}^t\!h_1
\,{}^t\!B^{-1}\hat{F}^{-1}h_1-2\,{}^t\!h_2B^{-1}(i+A)h_1-\\
&& \left. \left.  4\ii \,{}^t\!h_1\,{}^t\!B^{-1}
\hat{F}^{-1}\beta_1 \right] \right\}~.
\end{eqnarray}
In the following manipulations we will focus on the expression in the
square brackets only. It is useful to observe that the by redefining
\begin{equation}
  \gamma_1=\hat{F}^{-1}\beta_1 \,\,\,\,\,\, 
\mathrm{and} \,\,\,\,\,\, k=\hat{F}^{-1}h_1
\end{equation}
and making use again of the identities mentioned earlier, involving
the entries of the complex structure and of $\hat{F}$, one can rewrite the
content of the square brackets above as
\begin{eqnarray}
\nonumber && {}^t\!h_2B^{-1}(i+A)\hat{F}h_2+2\ii \,
{}^t\!\gamma_1B^{-1}\hat{F}\gamma_1+\,{}^t\!
h_2B^{-1}(i+A)\hat{F}\gamma_1+\\
\nonumber && +\,{}^t\!\gamma_1B^{-1}(i+A)
\hat{F}h_2+2\,{}^t\!\beta_2h_2+2\ii \,{}^t\!kB^{-1}\hat{F}k-\\
&& -2\,{}^t\!h_2B^{-1}(i+A)\hat{F}k-4\ii \,{}^t\!kB^{-1}\hat{F}\gamma_1 ~.
\end{eqnarray}
As $k$ in general is no longer a column of  integers, it is convenient
 to write it
distinguishing its integer part from the remainder
 \begin{equation}
k= r+\hat{F}^{-1}l ~,
\end{equation}
with $r \in \mathbb{Z}^d$ and $l_{\alpha} \in
[1,\hat{F}_{\alpha\alpha}]$. Thus the expression above reads
\begin{eqnarray}
  \nonumber && 2 \,{}^t\!\left( r+\hat{F}^{-1}l-\gamma_1 \right)
\ii B^{-1}\hat{F} \left( r+\hat{F}^{-1}l-\gamma_1 \right)+\,{}^t\!h_2
B^{-1}(i+A)\hat{F}h_2+\\
  && -2\,{}^t\!h_2B^{-1}(i+A)\hat{F}
\left( r+\hat{F}^{-1}l-\gamma_1 \right) +2\,{}^t\!\beta_2h_2~.
\end{eqnarray}
Finally, defining $s=r-h_2 \in \mathbb{Z}^d$, one has
\begin{eqnarray}
\nonumber {}^t\!\left( r+\hat{F}^{-1}l-\gamma_1 \right) 
B^{-1}(i-A)\hat{F} \left( r+\hat{F}^{-1}l-\gamma_1 \right) +
2\,{}^t\! \left( r+\hat{F}^{-1}l-\gamma_1 \right) \beta_2+\\
{}^t\!\left( s+\hat{F}^{-1}l-\gamma_1 \right)B^{-1}(i+A)\hat{F} 
\left( s+\hat{F}^{-1}l-\gamma_1 \right)-2\,{}^t\!
\left( s+\hat{F}^{-1}l-\gamma_1 \right) \beta_2~.
\end{eqnarray}
Thus the factorized amplitude reads
\begin{eqnarray}\label{FactAmpl}
\nonumber \mathcal{A}=
\sum_{l_{\alpha}=1}^{\hat{F}_{\alpha\alpha}} 
\frac{1}{\sqrt{\mathrm{Det}(2B\hat{F})}} && \vartheta \left[
\begin{array}{c}
\hat{F}^{-1}(l-\beta_1)\\
\beta_2
\end{array} \right] \left( 0 \big| B^{-1}(i-A)\hat{F} \right) \times\\
&& \vartheta \left[
\begin{array}{c}
\hat{F}^{-1}(l-\beta_1)\\
-\beta_2
\end{array} \right] \left( 0 \big| B^{-1}(i+A)\hat{F} \right)
\end{eqnarray}
written in terms of $d$-dimensional Theta-functions 
\begin{equation}
\vartheta \left[
\begin{array}{c}
a\\ b
\end{array} \right] \left(\nu | \tau \right)= 
\sum_{h \in \mathbb{Z}^d} \mathrm{exp}\left[ \pi \ii \,
{}^t\!(h+a)\tau(h+a)+2\pi \ii\,{}^t\!(\nu+b)(h+a) \right]~.
\end{equation}
Notice that the function in the second line of the
Eq.~(\ref{FactAmpl}) is indeed the complex conjugate of the one in the
first line, as $\hat{F}$, $A$ and $B$ are real $d \times d$ matrices
since $\mathcal{I}$ in Eq.~(\ref{RealComplexStructure}) is
real. This function is to be interpreted as the classical contribution
to the Yukawa couplings for three twisted states arising in a generic
$T^{2d}$ compactification of string theory with magnetised space
filling D-Branes. The sum in front of the couplings reveals their
multiplicity, given by $\mathrm{Det}\hat{F}=
\prod_{\alpha=1}^d\hat{F}_{\a\a}$. 
 Finally for the sake of
completeness let us write the expression for the correlator between
three twist fields (fixing one $l$, i.e. one particular coupling)
including also its quantum contribution in the $\mathcal{N}=1$
supersymmetric configuration
$\epsilon^\a_1+\epsilon^\a_2+\epsilon^\a_3=1$\footnote{We use the
  notations of Sect.4 of \cite{Russo:2007tc}.}:
\begin{eqnarray} \nonumber 
\langle \sigma_{\epsilon_1}\sigma_{\epsilon_2}
\sigma_{\epsilon_3}\rangle & = &  \prod_{i=1}^3
\prod_{\a=1}^d \left[ \frac{\Gamma(1-\epsilon^\a_i)}{\Gamma(\epsilon^\a_i)} 
\right]^{\frac{1}{4}}
\left( \mathrm{Det}(2B) \right)^{-\frac{1}{4}}\\
& \times & \vartheta\left[
\begin{array}{c}
\hat{F}^{-1}(l-\beta_1)\\
\beta_2
\end{array} \right] \left( 0 \big| B^{-1}(i-A)\hat{F} \right)~.
\end{eqnarray}
We can check that this result is in agreement with the literature
considering in particular the multiplicity of the Yukawa couplings in
the case of parallel fluxes, i.e. when all of the boundaries are
magnetized D-Branes with magnetic fields of the type (\ref{Fblock})
put in the form (\ref{Fblock1}). Notice that this setup can always be
T-dualized into a configuration of intersecting D-Branes on a
$2d$-dimensional torus that is not geometrically a direct product of
$d$ $T^2$'s. However in the counting of the non vanishing 3-point
correlators between twisted states the metric of the torus is not
involved, and thus we expect this multiplicity is equal to the one
already calculated~\cite{Cremades:2003qj} in fully factorizable models
of D-Branes at angles on $(T^2)^d$. Indeed following the steps of the
previous subsection it is not difficult to see that $H=w_1w_2/\delta$,
$\delta$ being a diagonal matrix whose eigenvalues are the G.C.D.'s of
the entries of $w_1$ and $w_2$ as in Eq.~(\ref{wma}). Then by
generalizing also the definition (\ref{Iij}) into a diagonal matrix,
with the same structure and the intersection numbers as entries, one
has $\hat{F}=I_{21}w_1w_2/\delta^2$. Upon the straightforward
generalization of the T-Duality in Eq.~(\ref{zeroDuality}) $\hat{F}
\rightarrow I_{21}I_{20}I_{01}/\delta^2$ where $\delta$ now contains
the G.C.D.'s of the entries of the three intersection numbers and
\begin{equation}
\mathrm{Det}\hat{F}=
\prod_{\alpha=1}^d\frac{(p_{2\alpha}W_{0\alpha}-p_{0\alpha}W_{2\alpha})
(p_{0\alpha}W_{1\alpha}-p_{1\alpha}W_{0\alpha})
(p_{2\alpha}W_{1\alpha}-p_{2\alpha}W_{2\alpha})}{\delta^2_{\alpha\alpha}}
\end{equation}
This number agrees with the product of the multiplicity of Yukawa
couplings in each of the $T^2$'s inside the $T^{2d}$ defined by the
form of the magnetic fields (see \cite{Cremades:2003qj}).

\vspace{1cm}

\noindent {\large {\bf Acknowledgements}}

\vspace{3mm}

\noindent 
We wish to thank Giulio Bonelli, Paolo Di Vecchia, Francisco Morales,
Igor Pesando, Sanjaye Ramgoolam, Alessandro Tanzini and Steve Thomas
for useful discussions and comments.  This work is partially supported
by the European Community's Human Potential under contract
MRTN-CT-2004-005104 and MRTN-CT-2004-512194, and by the Italian MIUR
under contract PRIN 2005023102. ~R. R. and S. Sc. thank the
Galileo Galilei Institute for Theoretical Physics, and S. Sc. thank the
Queen Mary University, for hospitality during the completion of this work.

\appendix

\sect{Generic and block-diagonal integer matrices}\label{AppendixFTransf}

A generic integer $2d \times 2d$ antisymmetric matrix $M$ can always
be put in a block-diagonal form by means of a transformation of the
type $M \rightarrow {}^tOMO$, $O$ being a unimodular integer matrix. 

In order to show that this is indeed the case one can observe that any
antisymmetric matrix:
\begin{equation}\label{M}
M=\left(
\begin{array}{cc}
A & B\\
-{}^tB & C
\end{array} \right)
\end{equation}
where
$A$ and $C$ are $2k \times 2k$ and  $2(d-k) \times 2(d-k)$
antisymmetric matrices and $B$ is a rectangular  $2k \times 2(d-k)$
matrix, can be block diagonalized by
\begin{equation}\label{O}
O=\left(
\begin{array}{cc}
1_{2k} & -A^{-1}B\\
0 & 1_{2(d-k)}
\end{array} \right),
\end{equation}
getting:
\begin{equation}\label{OMO}
{}^tOMO=\left(
\begin{array}{cc}
A & 0\\
0 & {}^tBA^{-1}B+C
\end{array} \right)
\end{equation}
Notice that each block of ${}^tOMO$ is also an antisymmetric matrix
and that the determinant of the matrix $O$ is one. Since the form of
${}^tOMO$ in Eq.~(\ref{OMO}) is independent of the choice of $k$, one
can use an iterative procedure to obtain the final block diagonal
form
always choosing $k=1$.
Using this procedure $d-1$ times one finds
\begin{equation}\label{M''}
{}^tO M O
=\left(
\begin{array}{cccccc}
a_1 & 0 & 0 & 0 & \cdots & \cdots\\
0 & a_2 & 0 & 0 & \cdots  &\cdots\\
\vdots & \vdots & \vdots & \vdots &  \vdots &  \vdots\\
0 & 0 & 0 & 0 &  \cdots & a_d \\
\end{array} \right) \otimes 
\left(
\begin{array}{cc}
0 & 1 \\
-1 & 0 
\end{array} \right),
\end{equation}
where: $O=O_1 O_2...O_{d-1}$.
By a suitable permutation of the rows and of the columns of the matrix above
one can rewrite the transformed matrix as
\begin{equation}\label{M'}
M'= \left(
\begin{array}{cc}
0 & \tilde{M}\\
-\tilde{M} & 0
\end{array} \right),
\end{equation}
where $\tilde{M}=\mathrm{diag}\{a_1,a_2, \dots \,a_d \}$.

If the matrix $M$ has integer elements, the transformed matrix will be
integer only if $\mathrm{Det}A=1$.  We will now show that it is always
possible to reduce to this case at any step of the iterative
procedure. Obviously, if at least one element of the matrix $M$ is
$1$, it is enough to suitably relabel the rows and the columns.
Otherwise, if $\mathrm{Det}A \neq 1$ we can distinguish various cases.
The simplest one is when in a row two elements are coprime. By
relabeling of the rows and the columns it is possible to put these
elements ($a$ and $b$) in the first row as follows:
\begin{equation}\label{M2}
M
=\left(
\begin{array}{cccc}
0 & a & b & \cdots\\
-a & 0 & c & \cdots\\
-b & -c & 0 & \cdots\\
\vdots & \vdots & \vdots & \ddots
\end{array} \right).
\end{equation}
Then it is easy
to see that the unimodular integer matrix
\begin{equation}\label{O1}
Q=\left(
\begin{array}{ccccc}
1 & 0 & 0 & 0 & \cdots\\
0 & x & -b & 0 & \cdots\\
0 & y & a & 0 & \cdots\\
0 & 0 & 0 & 1 & \cdots\\
\vdots & \vdots & \vdots & \vdots & \ddots
\end{array} \right),
\end{equation}
with $x$, $y$  solution of the Diophantine equation $ax+by=1$,
transforms $M$ into:
\begin{equation}
M \rightarrow {}^t Q M Q = \left(
\begin{array}{cccc}
0 & 1 & 0 & \cdots\\
-1 & 0 & c & \cdots\\
0 & -c & 0 & \cdots\\
\vdots & \vdots & \vdots & \ddots
\end{array} \right),
\end{equation}
which has $\mathrm{Det}A=1$.

If, instead, there is no row with two coprime elements, but in
Eq.~\eq{M2} $a\neq \pm b$ and their greatest common denominator $d$ is
different from $1$, one can repeat the previous step with the matrix:
\begin{equation}\label{O2}
Q'=\left(
\begin{array}{ccccc}
1 & 0 & 0 & 0 & \cdots\\
0 & x & -\frac{b}{d} & 0 & \cdots\\
0 & y & \frac{a}{d} & 0 & \cdots\\
0 & 0 & 0 & 1 & \cdots\\
\vdots & \vdots & \vdots & \vdots & \ddots 
\end{array} \right),
\end{equation}
where now $x$ and $y$ solve the Diophantine equation $ax+by=d$.
This yields
\begin{equation}
M \rightarrow \,{}^t\!Q'MQ'=\left(
\begin{array}{cccc}
0 & d & 0 & \cdots\\
-d & 0 & c & \cdots\\
0 & -c & 0 & \cdots\\
\vdots & \vdots & \vdots & \ddots
\end{array} \right).
\end{equation}
One has to apply this procedure (which does not change the other
elements of the first row, from the 4th column on) till the matrix is
reduced to one of the following cases: either in the first row of the
transformed $M$ there are two coprime non-vanishing entries, then one
can use the matrix $Q$ to obtain $a_1=1$; or all of the non-zero
elements there coincide with $\pm d$. This is the case if for instance
in the original matrix $M$ the first row contained elements which were
all multiples of $d$.  If there is any other row in the transformed
matrix with two different non-zero elements $a \neq \pm b$ for which
$d$ is not a divisor, then, by exchanging rows and columns among
themselves, it is possible to bring this as the first row and reapply
the transformations encoded in $Q$ or $Q'$. Otherwise one can have two
possible forms for the transformed matrix.  One possibility is that
all of the elements of $M$ are integer multiples of $d$. In this case
the common divisor $d$ can be factored out to reduce to the case with
$\mathrm{Det} A=1$.

 The other possibility is that the matrix has diagonal blocks, in
 which all the elements are multiple of different integers $d_i$. 
If all the non trivial blocks are $2\times 2$, we have got our aim; otherwise
 we can factor out  $d_i$ from the block of larger
 dimensions; for each of them we are again reduced to the case with 
 $\mathrm{Det} A=1$. Repeating the procedure, if needed, we finally
 end to the matrix: 
\begin{equation}
M_1=\left(
\begin{array}{ccccc}
0 & d_1 & 0 & 0 & \cdots\\
-d_1 & 0 & 0 & 0 & \cdots\\
0 & 0 & 0 & d_2 & \cdots\\
0 & 0 & -d_2 & 0 & \cdots\\
\vdots & \vdots & \vdots & \vdots & \ddots
\end{array} \right)
\end{equation}
with all integer elements.

Although this already is the final form we are after, for sake of
completeness we recall that the normal form of the initial
matrix~(\ref{M''}), as discussed in \cite{Griffiths}, has the further
property that $a_{\a+1}/a_\a \in \mathbb{N}$, $\forall\a$. In order to
achieve this (even if it is not strictly necessary for the
computations considered here) one can use the following transformation
\begin{equation}
Q''=\left(
\begin{array}{ccccc}
1 & 0 & 0 & 0 & \cdots\\
0 & 1 & 0 & 0 & \cdots\\
1 & 0 & 0 & 1 & \cdots\\
0 & 0 & 1 & 0 & \cdots\\
\vdots & \vdots & \vdots & \vdots & \ddots
\end{array} \right)
\end{equation}
that mixes the $d_i$'s and gives back a form that can be reduced by
means of either $Q$ or $Q'$. Following this procedure, it is possible
to convince oneself that the final matrix~(\ref{M''}) entries satisfy
the property mentioned above, since one actually ends the repeated
application of $Q$, $Q'$, and $Q''$ only if in the first $2 \times 2$
block there is a one, or if the matrix is proportional to an integer
as a whole.

Coming back to our main problem, we can now apply the procedure
outlined in this Appendix to rewrite the quantized magnetic
field~(\ref{Fquantization}) on a generic magnetized D-Brane in the
form~(\ref{Fblock}).  It is first convenient to define an integer
matrix $P$ associated to the magnetic field~(\ref{Fquantization})
\begin{equation}
\label{p}
  P=F \times \mathrm{m.c.m}\left\{ w_M w_N,\, M\neq N,~~\,\forall
  M,N=1, \dots,2d
  \right\}
= \omega F.
\end{equation}
where $\omega$ is the minimum common multiple of all the pairs of
wrappings that appear in the denominators of
Eq.~(\ref{Fquantization}).  The integer matrix $P $ can now be
transformed into a block-diagonal form as in Eq.~(\ref{M''}) by means
of an integer unimodular matrix $O$ that preserves the lattice of the
torus as
\begin{equation}
\label{cell}
P \rightarrow \,{}^t\!OPO=
\omega\,{}^t\!OFO=\omega F_{\mathrm{block}}.
\end{equation}
Hence $F_{\mathrm{block}}$ will have the same form as Eq.~(\ref{M''}) with
rational entries whose numerators and denominators could still have
factors in common. By expurgating these factors one exactly recovers
the form in Eq.~(\ref{Fblock}).
First of all we will show that the phase factor in the boundary state~\eq{WLBF}
is not affected by the change of the fundamental cell in the lattice
torus performed  in~\eq{cell}. It reads:
\begin{equation}\label{ph}
\mathrm{Ph}=\mathrm{exp}\left[\ii \pi\sum_{M<N}
\hat{m}^MF_{MN}\hat{m}^N \right]
=\mathrm{exp}\left[\frac{ \ii \pi}{\omega}\sum_{M<N}
\hat{m}^MP_{MN}\hat{m}^N \right]
\end{equation}
where $P$ is the integer matrix defined in~\eq{p}, that we write in
the form of Eq.~\eq{M}.
To write it as a block diagonal matrix, we use the techniques just discussed; focusing at first on the simplest case with
\begin{equation}\label{A}
A =\left(
\begin{array}{cc}
0 & 1\\
-1 & 0
\end{array} \right),
\end{equation}
let us consider how the phase $\mathrm{Ph}$ of~\eq{ph} transforms under the substitution:  
$\hat{m}=O \hat{m}'$, with
$O$ as in Eq.~(\ref{O}).
One gets:
\begin{eqnarray}\label{ph'}
  \nonumber \mathrm{Ph}=\mathrm{exp} \left[ \frac{\ii \pi}{\omega}\left(
      \hat{m}'_1 A_{12}\hat{m}'_2+
      \sum_{j=3}^{2d}B_{2j}\hat{m}'_jA_{12}\hat{m}'_2 
      -\hat{m}'_1A_{12}\sum_{j=3}^{2d}B_{1j}\hat{m}'_j 
    \right. \right.\\
  \nonumber -\sum_{j,k=3}^{2d}B_{2j}\hat{m}'_jA_{12}B_{1k}\hat{m}'_k+
  \sum_{j=3}^{2d}\hat{m}'_1B_{1j}\hat{m}'_j+
  \sum_{j,k=3}^{2d}B_{2j}\hat{m}'_jB_{1k}\hat{m}'_k\\ 
  \left. \left. +\sum_{j=3}^{2d}\hat{m}'_2B_{2j}\hat{m}'_j-
      \sum_{j,k=3}^{2d}B_{1j}\hat{m}'_jB_{2k}\hat{m}'_k +
      \sum_{k>j=3}^{2d}\hat{m}'_jC_{jk}\hat{m}'_k \right) \right]
\end{eqnarray}
By remembering that all the winding numbers $\hat{m}^N$ must be
integer multiples of the corresponding wrapping numbers $w_N$, one can
check that, in spite of the denominator $\omega$, all the terms in the
exponent are integer multiples of $\ii \pi$. In fact combining the
form of the matrices~(\ref{O}) and~(\ref{p}), one finds the following
expressions for $\hat{m}=O\hat{m}'$
\begin{eqnarray}
\nonumber \hat{m}_1 & = & \hat{m}'_1+\frac{\omega}{w_2}
\sum_{i=3}^{2d}p_{2i}\frac{\hat{m}'_i}{w_i}\\
\nonumber \hat{m}_2 & = & \hat{m}'_2-\frac{\omega}{w_1}
\sum_{i=3}^{2d}p_{1i}\hat{m}'_i\\
\hat{m}'_i & = & \hat{m}_i
\end{eqnarray} 
In this case, in order to have the matrix $A$ in the form~(\ref{A}),
it is necessary 
that $p_{12}=1$ and $\omega=w_1w_2$, hence, since from the last line
of the previous equation $\hat{m}'_i/w_i$ must be integer, it is also
true, in the first and second line, that the transformed winding
numbers $\hat{m}^{'N}$ are integer multiples of the wrapping numbers
$w_N,~ \forall N=1,2,...,2d$. 
Thus each of the terms in the sum~(\ref{ph'}) is an integer number, as
one can check for instance considering the first term of the second
line in Eq.~(\ref{ph'}); writing $\hat{m}^{'N}=m^{'N}w_N$ with $m^{'N}
\in \mathbb{Z}$ one gets:
\begin{equation}
\frac{1}{\omega}B_{2j}\hat{m}'_jA_{12}B_{1k}\hat{m}'_k=
\frac{1}{\omega} \frac{\omega p_{2j}}{w_2w_j}m'_jw_j \frac{\omega
  p_{1k}}{w_1w_k}m'_kw_k=p_{2j}m'_jp_{1k}m'_k \in \mathbb{Z}. 
\end{equation} 
So we can freely change the sign of each term in Eq.~(\ref{ph'}), obtaining
\begin{eqnarray}\label{phf}
\mathrm{Ph} & = & \mathrm{exp} \left[ \ii \pi \left(
    \sum_{M<N} \hat{m}^{'M}({}^tO FO)_{MN}
\hat{m}^{'N}+\frac{1}{\omega}
\sum_{j=3}^{2d}B_{1j}B_{2j} \hat{m}_j^{'2} \right) \right]=\\ 
& = & \mathrm{exp} \left[ \ii \pi \left( \sum_{M<N} \hat{m}^{'M}({}^tO
  FO)_{MN}
\hat{m}^{'N}+
\sum_{j=3}^{2d}p_{1j}p_{2j} \frac{\hat{m}'_j}{w_j} \right) \right],
\end{eqnarray}
where we have used the explicit expression of $B_{1j}$ and $B_{2j}$ in
terms of the Chern numbers $p_{1j}$ and $p_{2j}$ and of the winding
numbers; moreover we have taken into account the fact that $(m'_j)^2$
has the same parity (even/odd) as $m'_j=\hat{m}'_j/w_j$. Thus the
phase factor can be written in terms of the transformed field ${}^tO
FO$ and of the transformed winding numbers $\hat{m}'$'s with the same
functional form as the original one~(\ref{ph}), with a half-integer
shift of the Wilson line when $p_{1j}p_{2j}$ is odd.

If ${}^tO FO$ is already block diagonal, we have ended our job,
otherwise we have to repeat the procedure. In an analogous fashion, if
the entries of $A$ are not equal to one, one can check that the
transformations related to the matrices in Eq.~(\ref{O1})
and~(\ref{O2}), involved in reducing $A$ to the form considered in the
previous example, also preserve the form of the phase factor up to
half-integer Wilson lines. With similar manipulations it is also
possible to prove, in a basis in which $(F_2-F_1)$ is block-diagonal,
that the phase factors in Eq.~\eq{Azm2} follow from those in
Eq.~(\ref{e2finalg2}). As usual, one has to introduce $h \in
\mathbb{Z}^{2d}$ by using $\hat{m}_1=Hh$; then it is possible to check
that the combination $\sum_{M<N}(Hh)^M (F_2-F_1)_{MN} (Hh)^N$ is
equal, modulus two, to $\sum_{M<N} h^M\left[H(F_2-F_1)H\right]_{MN}
h^N$, apart from terms quadratic in $h^M$ that can be reabsorbed into
a half-integer shift of the Wilson lines.

Finally we mention that the transformations discussed in this Appendix
do not affect the other contributions to the amplitude, in the
effective field theory limit that we consider for the factorization,
if one suitably redefines the complex structure in
Eq.~(\ref{e2finalg2}).  Indeed the combination
${}^t\hat{m}_1\mathcal{I}F\hat{m}_1$ can be rewritten, by redefining
$\hat{m}_1=O\hat{m}'_1$, as
\begin{equation}
{}^t\!\hat{m}'_1\,{}^t\!O\mathcal{I}FO\hat{m}'_1=
 {}^t\hat{m}'_1\,{}^t\!O\mathcal{I}{}^t\!O^{-1}\,{}^t\!OFO\hat{m}'_1 
=\,{}^t\!\hat{m}'_1\mathcal{\hat{I}}F_{\mathrm{block}}\hat{m}'_1,
\end{equation}
$\mathcal{\hat{I}}$ still being a good complex structure and
 $F_{\mathrm{block}}$
being in the form~(\ref{M''}).

\end{document}